%% file: 0main.tex
\documentclass[sigconf,nonacm]{acmart}
\AtBeginDocument{%
  }

\usepackage{graphicx}
\usepackage{textcomp}
\usepackage[table,x11names,dvipsnames,table,xcdraw]{xcolor}

\usepackage{framed}
\usepackage{hyperref}
\usepackage{booktabs}
\usepackage{multirow}
\usepackage{algorithm}
\usepackage{algpseudocode}
\usepackage{circledsteps}
\usepackage{tabularx}
\usepackage{needspace}

\usepackage[utf8]{inputenc}
\newcommand{\F}{\mathbb{F}}
\newcommand{\HH}{\mathbb{H}}
\newcommand{\eq}{\mathsf{eq}}

\definecolor{rowgray}{gray}{0.95}
\definecolor{takeawaybg}{gray}{0.96}
\newenvironment{takeawaybox}
  {\Needspace{6\baselineskip}%
   \setlength{\OuterFrameSep}{0pt}%
   \MakeFramed{\advance\hsize-\width \FrameRestore}%
   \noindent\textbf{Key Takeaway.}\ }
  {\endMakeFramed}

\begin{document}

\title[]{When Proofs Meet Hardware: Comparing NTT and SumCheck in Zero-Knowledge Systems}

\author{Jianqiao Mo}
\authornote{\small{These authors contributed equally to this work.}}
\authornote{\small{This work was partially done during a research internship at Intel.}}
\affiliation{%
  \institution{New York University, Intel}
  \city{Brooklyn}
  \state{NY}
  \country{USA}
}
\email{jm8782@nyu.edu}

\author{Alhad Daftardar}
\authornotemark[1] % References the first authornote (Equal Contribution)
\affiliation{%
  \institution{New York University}
  \city{Brooklyn}
  \state{NY}
  \country{USA}
}
\email{ajd9396@nyu.edu}

\author{Barath GaneshKumar}
\affiliation{
    \institution{New York University}
    \city{Brooklyn}
    \state{NY}
    \country{USA}
}
\email{bg2697@nyu.edu}

\author{Kaiyue Guo}
\affiliation{
    \institution{New York University}
    \city{Brooklyn}
    \state{NY}
    \country{USA}
}
\email{kg3209@nyu.edu}

\author{Hong Wang}
\affiliation{
    \institution{Intel}
    \city{Santa Clara}
    \state{CA}
    \country{USA}
}
\email{hong_wang@ieee.org}

\author{Benedikt B{\"u}nz}
\affiliation{
    \institution{NYU Courant}
    \city{New York}
    \state{NY}
    \country{USA}
}
\email{bb@nyu.edu}

\author{Siddharth Garg}
\affiliation{
    \institution{New York University}
    \city{Brooklyn}
    \state{NY}
    \country{USA}
}
\email{sg175@nyu.edu}

\author{Brandon Reagen}
\affiliation{
    \institution{New York University}
    \city{Brooklyn}
    \state{NY}
    \country{USA}
}
\email{bjr5@nyu.edu}

\renewcommand{\shortauthors}{}

\begin{abstract}

In the ZKP community, it has long been discussed that the {SumCheck} protocol is asymptotically more efficient than the {Number Theoretic Transform} (NTT), requiring only $O(N)$ arithmetic versus $O(N \log N)$. At the same time, hardware accelerator designers propose that NTT is more hardware-friendly, benefiting from locality and data reuse, while SumCheck suffers from sequential, dependent rounds.
Despite these competing intuitions, the hardware-system-level trade-offs between NTT- and SumCheck-based proving primitives remain insufficiently understood.

Beyond individual accelerator design, this work presents, to our knowledge, the first hardware-system-level direct comparison of NTT- and SumCheck-based proving primitives under a unified architectural framework.
We study them in the context of the ZeroCheck protocol, a common building block in zkSNARKs.
We implement optimized systems for both primitives.
Both are evaluated under the same level on-chip SRAM and off-chip bandwidth budgets.
Our results show that there is no universal winner. 
Generally, SumCheck outperforms NTT for high-degree polynomials.
For low-degree polynomials, performance depends on memory availability: 
under given SRAM budgets, NTT might deliver better performance for medium-sized workloads by exploiting data reuse. 

These findings, bridging cryptographic protocol design and hardware architecture, offer practical guidance for understanding the proving cost of NTT- and SumCheck-based zero-knowledge proof systems.

\end{abstract}

%    <concept>
%        <concept_id>10010520.10010521.10010542.10011714</concept_id>
%        <concept_desc>Computer systems organization~Special purpose systems</concept_desc>
%        <concept_significance>500</concept_significance>
%        </concept>
%    <concept>
%        <concept_id>10002978.10002979</concept_id>
%        <concept_desc>Security and privacy~Cryptography</concept_desc>
%        <concept_significance>500</concept_significance>
%        </concept>
%  </ccs2012>

% \ccsdesc[500]{Computer systems organization~Special purpose systems}
% \ccsdesc[500]{Security and privacy~Cryptography}

% \keywords{Zero-Knowledge Proof,
% hardware architecture}

\maketitle

\input{01Introduction}

\input{02Preliminaries}
\input{03Situations}

\input{04Arch}
\input{05Evaluation}
\input{06Related}

\begin{acks}
This work was supported in part by the NSF CAREER Award (NSF \#2340137), NSF CIRC GRAND: Cryptolets (NSF \#2450539), and generous support from DTCC, Intel and Google. 
Any opinions, findings, conclusions, or recommendations expressed in this material are those of the authors and do not necessarily reflect the views of NSF, DTCC, Intel, Google, or any other supporting organizations.
\end{acks}

\bibliographystyle{ACM-Reference-Format}
\bibliography{ref-sample-base}

\appendix

\end{document}

%% file: 01Introduction.tex
\section{Introduction}
\label{sec:intro}

Zero-knowledge proofs (ZKPs) \cite{zk_proof_gmr} allow a prover to convince a verifier of the correctness of a computation without revealing private inputs.
Modern zkSNARK constructions \cite{zksnarks_bcct12} typically combine a Polynomial Interactive Oracle Proof (PIOP) \cite{iops_bcs, hyperplonk} with a polynomial commitment scheme \cite{kzg_pcs}, using the Fiat-Shamir heuristic \cite{fiat1986prove} to remove interaction.

\begin{figure}[t!]
\centering
\vspace{1em}
\includegraphics[width=1\columnwidth]{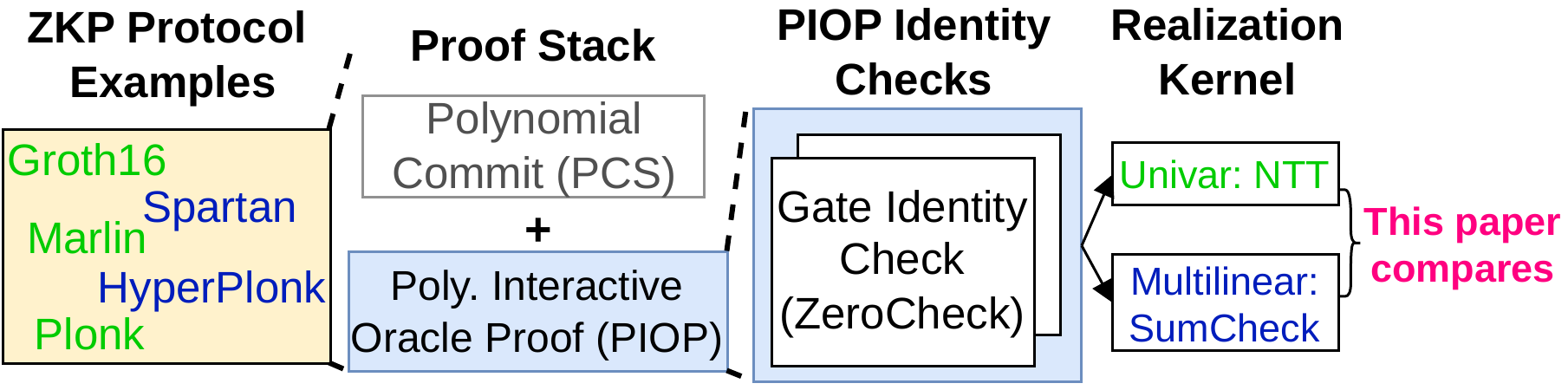}
\caption{ZKP protocols include several identity checks for proving different aspects of computation, realized either by univariate polynomials (NTT) or multilinear polynomials (SumCheck kernels). 
Both achieve the same checking goal.}
\label{fig:zerocheck_in_zkp}
\end{figure}

Two major PIOP families dominate current zkSNARK designs.
The first relies on \emph{univariate polynomials} and fast evaluation via the Number Theoretic Transform (NTT), as in Groth16 \cite{groth16} and Plonk \cite{gabizon2019plonk}.
The second is based on \emph{multilinear polynomials}, where the SumCheck protocol is the central primitive, as used in Aurora \cite{ben2019aurora}, HyperPlonk \cite{hyperplonk}, and Spartan \cite{spartan}.
Both families ultimately commit to polynomials using schemes such as KZG \cite{kzg_pcs} or Merkle-based commitments \cite{ligero_pcs}, but differ fundamentally in how the PIOP layer is structured.

From an asymptotic perspective, these approaches present different trade-offs.
NTT-based protocols require $O(N \log N)$ arithmetic but produce constant-size proofs, while SumCheck-based protocols achieve $O(N)$ arithmetic with proof sizes of $O(\log N)$.
This has led to the conventional view that SumCheck is asymptotically superior and potentially faster for large-scale computations.
In contrast, ``community wisdom'' of hardware intuition favors NTT-based designs.
SumCheck proceeds through sequential, round-dependent computation, limiting cross-round parallelism and throughput on accelerators (e.g., Ingonyama's discussion \cite{ingonyama_2022} of HyperPlonk).
NTT and FFT-based workloads, by comparison, map naturally to deeply pipelined and parallel hardware architectures (e.g., see \cite{irreducible_2023, janeStreet_2022}).
As a result, SumCheck is often viewed as hardware-unfriendly despite its asymptotic efficiency, while NTT is considered better aligned with accelerator design.
However, this belief remains difficult to assess without a principled, head-to-head hardware-system-level evaluation.

A direct comparison presents two key challenges.
First, comparing full prover implementations risks confounding effects from unrelated components such as commitment schemes or batching strategies.
Second, fairness requires optimized implementations of both primitives; while NTT accelerators are well studied, optimized SumCheck hardware remains relatively unexplored due to its iterative structure and memory demands.
To draw meaningful conclusions, both systems must be engineered to efficiently utilize modular multipliers and memory bandwidth.

In this work, we focus on the \emph{ZeroCheck} primitive, which verifies that a polynomial evaluates to zero over a given domain.
ZeroCheck is widely used in modern ZKP systems and can be instantiated using either univariate (NTT-based) or multilinear (SumCheck-based) PIOPs.
Both variants operate on the same input size and support similar commitment and batching techniques, making ZeroCheck a clean and fair vehicle for isolating PIOP-level hardware costs.
In the univariate setting, ZeroCheck relies on NTT and inverse NTT operations to derive a quotient polynomial, whereas in the multilinear setting it executes multiple rounds of SumCheck over multilinear extensions.

\textbf{Our Contributions.}
This work makes the following contributions:
\begin{itemize}
    \item Beyond an accelerator design, we present, to our knowledge, the first hardware-system-level direct comparison of SumCheck- and NTT-based proving primitives using ZeroCheck as a common benchmark, providing practical guidance for protocol and accelerator designers.
    \item We incorporate state-of-the-art optimizations for both systems, evaluating programmable SumCheck architectures alongside streaming and four-step decomposed NTT designs under identical resource constraints.
\end{itemize}

Our results show that there is no universal winner.
For medium-sized, low-degree workloads under constrained SRAM, NTT-based designs often perform better, particularly when polynomials contain many unique terms.
However, SumCheck demonstrates superior efficiency for high-degree workloads, where the computational cost of NTT scales rapidly with degree.
These findings highlight that while NTT benefits from strong locality and reuse, its asymptotic scaling cannot be fully mitigated by hardware alone.
All evaluations are conducted under identical on-chip SRAM and off-chip bandwidth constraints. 
Our simulation framework is available.\footnote{\href{https://github.com/beekayg15/zerocheck}{https://github.com/beekayg15/zerocheck}}

%% file: 02Preliminaries.tex
\section{Background}
\label{sec:prelim}

Zero-Knowledge Proofs are cryptographic protocols that allow a prover to convince a verifier that a computation was performed correctly, without revealing anything about the computation’s private inputs.
They serve as a foundational tool for privacy-preserving applications, ranging from authentication and blockchain transactions to verifiable cloud computation and privacy-preserving machine learning \cite{zkml, zkgpt, zkllm, zkcnn, dhyani2024privit, mo2023towards, geng2026ppimce,karthik,cho2024apint}.
The practicality of a ZKP protocol depends on multiple factors, including prover and verifier runtime, proof size, setup assumptions, and cryptographic security guarantees.
Different applications prioritize different trade-offs-for example, blockchain systems emphasize small proof sizes to reduce on-chain costs, whereas outsourced computation frameworks may emphasize prover efficiency.

\subsection{zkSNARKs}
Among the most widely used constructions are zkSNARKs (Zero-Knowledge Succinct Non-interactive Arguments of Knowledge).
zkSNARKs are built by combining a Polynomial Interactive Oracle Proof (PIOP) with a Polynomial Commitment Scheme (PCS).
In this framework, the PIOP encodes the computation into polynomial relations that the verifier can check probabilistically, while the PCS ensures that the committed polynomials are both binding and hiding.
The Fiat-Shamir heuristic is then applied to transform the interactive PIOP into a non-interactive protocol.
This modular design has enabled the development of numerous protocols with varying efficiency and security properties.

\autoref{fig:zerocheck_in_zkp} situates the scope of this paper within this stack.
The PIOP layer typically contains several polynomial identity checks, which prove different parts of the encoded computation: for example, that circuit gates are evaluated correctly, that copy or wiring constraints are satisfied, and that auxiliary consistency conditions hold.
Many of these checks (gate checks, wiring checks, and consistency checks) reduce to ZeroCheck-style claims that a polynomial vanishes over a prescribed domain.
Depending on the polynomial representation, these identity checks lead to two protocol families: univariate protocols, such as Plonk~\cite{gabizon2019plonk} and Marlin~\cite{chiesa2020marlin}, that realize the checks with NTT/iNTT-based kernels, and multilinear protocols, such as Spartan and HyperPlonk, that realize the checks with SumCheck-based kernels.
This realization layer is the comparison point of this paper.

\textbf{SumCheck}. At the core of modern PIOPs lie two fundamental proof gadgets: the \emph{SumCheck} protocol and the \emph{Number Theoretic Transform} (NTT).
Both are widely adopted as compute kernels in different polynomial IOP schemes.
The SumCheck protocol \cite{sumcheck} is central to protocols such as {Spartan}~\cite{spartan}, {Aurora}~\cite{ben2019aurora}, HyperPlonk \cite{hyperplonk}, and several interactive proof systems where multilinear extensions are the natural representation of the computation.
It provides a way to prove that a multivariate polynomial evaluates to a ``claimed sum'' over the Boolean hypercube by progressively reducing the polynomial's dimension through rounds of interaction.
Its efficiency stems from low-degree arithmetic per round, but it requires repeated updates of large multilinear extension (MLE) tables, which can introduce significant bandwidth demands in hardware implementations.
Hardware accelerators like zkSpeed \cite{zkspeed2025}, NoCap \cite{nocap} target SumCheck as the core component.  

\begin{figure*}
\centering
\begin{framed}
\raggedright
\begin{minipage}[t]{0.48\textwidth}
\textbf{Univariate ZeroCheck (NTT-based)}
\begin{algorithmic}[1]
\State \textbf{Goal:} Prove $f:=g_1 g_2 + g_3$ vanishes on $\HH$ with $|\HH|=N=2^n$ (i.e.\ $z_{\HH} \mid f$)
\State \textbf{Input:} Evaluations of $g_1,g_2,g_3$ over $\HH$
\State $g_1,g_2,g_3 \leftarrow \text{INTT}_{\HH}(\text{evals}_{g_1},\text{evals}_{g_2},\text{evals}_{g_3})$
\State Pick a multiplicative coset $D$ with $|D|=\Theta(N)$ and where $z_{\HH}$ has no zeros
\State $(\text{evals}_{\hat g_1},\text{evals}_{\hat g_2},\text{evals}_{\hat g_3}) \leftarrow \text{NTT}_D(g_1,g_2,g_3)$
\State $\text{evals}_{\hat z} \leftarrow \text{Evaluate}_D(z_{\HH})$
\State $\text{evals}_{\hat f} \leftarrow \text{evals}_{\hat g_1} \cdot \text{evals}_{\hat g_2} + \text{evals}_{\hat g_3}$ (\textit{for each evaluation})
\State $\text{evals}_{q} \leftarrow \text{evals}_{\hat f} \oslash \text{evals}_{\hat z}$ \textit{(element-wise division)}
\State $q \leftarrow \text{INTT}_D(\text{evals}_{q})$
\State Commit: $[g_1],[g_2],[g_3],[q]$
\State $r \gets \mathsf{Hash}([g_1],[g_2],[g_3],[q])$
\State Open $g_1(r), g_2(r), g_3(r), q(r)$
\end{algorithmic}
\end{minipage}\hfill
\begin{minipage}[t]{0.48\textwidth}
\textbf{Multilinear ZeroCheck (SumCheck-based)}
\begin{algorithmic}[1]
\State \textbf{Goal:} 
Prove $f_M:=\tilde g_1\tilde g_2+\tilde g_3$ is $0$ on $\{0,1\}^{n}$ 
\State \textbf{Input:} MLEs $\tilde g_1,\tilde g_2,\tilde g_3$ over $\{0,1\}^{n}$ (committed)
\State $r \gets \mathsf{Hash}([\tilde g_1],[\tilde g_2],[\tilde g_3])$
\State Build MLE: 
$\eq_r(x)=\prod_{i=1}^{n}\big((1-x_i)(1-r_i)+x_i r_i\big)$
\State For writing simplicity, let 
$\hat{f}(x)=f_M(x)\,\eq_r(x)$
\State Claim $\sum_{x\in\{0,1\}^{n}}\hat{f}(x) =0$
\For{$i=1$ to $n$}
  \State Prover sends
\[
\begin{aligned}
G_i(X_i)
&= \sum_{x_{i+1},\ldots,x_n \in \{0,1\}}
   \hat{f}(\alpha_1,\ldots,\alpha_{i-1},X_i, x_{i+1},\\
&\qquad\qquad\qquad\qquad\qquad \ldots,x_n)
\end{aligned}
\]
  \State Set next challenge $\alpha_i \gets \mathsf{Hash}(G_i)$
\EndFor
\State Open $\tilde g_1,\tilde g_2,\tilde g_3$ at $\alpha=(\alpha_1,\ldots,\alpha_n)$
\end{algorithmic}
\end{minipage}
\end{framed}
\vspace{-1em}
\caption{Two types of ZeroCheck: Univariate approach uses INTT of inputs, NTT on a size-$\Theta(N)$ coset, element-wise division by $z_{\HH}$, and INTT to recover $q$. 
Multilinear approach includes SumCheck with kernel $\eq_r$, folding over $n=\log_2 N$ rounds with Fiat-Shamir challenges $\alpha_i$.}
\label{fig:side-by-side-zc}
\end{figure*}

\textbf{NTT}. In contrast, univariate polynomial IOPs typically rely on the Number Theoretic Transform (NTT), which is a key primitive in protocols such as {Groth16}~\cite{groth16} and {Plonk}~\cite{gabizon2019plonk}.
The NTT is essential in this setting because univariate protocols frequently require efficient conversion between a polynomial’s coefficient form (needed for commitments) and its evaluation form over a subgroup domain (for quotient computation).
Without the NTT, this transformation would require quadratic time, rendering large-scale proofs impractical.
By enabling $O(N \log N)$ conversions, the NTT makes these protocols efficient in practice.
Furthermore, NTT-based approaches benefit from data reuse, since loaded data can be repeatedly applied across multiple butterfly operations until the multi-round transform completes.
As a result, NTTs are generally considered compute-intensive, 
making them an attractive target for hardware acceleration.

A rich line of research has proposed optimizations for NTT to further improve efficiency \cite{franchetti2011fft}.
These include:  
\emph{constant-geometry} NTT \cite{pease1968adaptation}, which arranges butterflies in a highly regular pattern to facilitate pipelined hardware, reducing the complexity of data movement; 
\emph{four-step} NTT \cite{fourstep}, which decomposes a large transform into smaller mini-NTTs to better match small memory capacity;  
\emph{step-radix} NTT \cite{libfqfft}, which adapts the radix decomposition dynamically to handle sizes that are not exact powers of two; and  
\emph{sparse} NTT techniques, which exploit the sparsity of polynomial coefficients to avoid unnecessary computation and memory traffic.
Together, these optimizations enable practical deployment of NTT-based protocols across diverse hardware platforms.

Throughout this paper we denote:
a univariate polynomial $f(x)$ of degree $N = 2^n$ corresponds to a dataset of size $N$, i.e., a lookup table with $N$ entries. 
Similarly, a multilinear polynomial $f(x_1, \ldots, x_n)$ also represents size-$N$ data. 
In typical zero-knowledge proof systems, $n$ ranges between 16 and 32. 
When it is necessary to distinguish between the two settings, we write $f$ for univariate polynomials and $\tilde{f}$ for multilinear polynomials.

%% file: 03Situations.tex
\section{ZeroCheck Dataflow}
\label{sec:zerocheck}

\subsection{ZeroCheck in the ZKP Proof Stack}

ZeroCheck is a common way to express polynomial identity checks in modern ZKP systems.
A ZKP frontend first reduces the statement to an algebraic circuit or constraint system.
The prover must then show that the witness satisfies all constraints, such as gate equations, copy or wiring constraints, and auxiliary consistency checks.
In polynomial IOPs, these constraints are encoded as polynomial identities.
For example, Plonkish protocols include gate and permutation identities \cite{gabizon2019plonk}, while Groth16-style quadratic arithmetic checks that a polynomial such as $P(X)=A(X)B(X)-C(X)$ vanishes on the evaluation domain.
At a high level, these checks ask the prover to show that a polynomial evaluates to 0 at every point in a specified domain.

\autoref{fig:zerocheck_in_zkp} shows the main modules involved in this part of a ZKP proof stack: the PCS, the PIOP identity-check layer, and the realization kernel used inside that layer.
Operationally, a ZKP prover workflow \textit{(1)} first commits to witness polynomials, which binds the prover to fixed proving material (with PCS).
It then \textit{(2)} runs the PIOP identity checks (ZeroCheck), and \textit{(3)} opens the committed polynomials at verifier-derived challenge points (with PCS).
If the identity-check relation and the opening results agree, the verifier accepts that the proof is consistent and valid.
Our focus is the middle stage.
That identity-check stage can be realized in two different polynomial representations: 
in univariate protocols, the prover realizes the identity check through NTT-based quotient computation; in multilinear protocols, the prover reduces the same vanishing claim through SumCheck.
Both paths serve the same logical role in the proof, but they induce different data movement, arithmetic intensity, and parallelism.

This paper therefore uses ZeroCheck as the unit of comparison.
By holding the high-level proof goal fixed and changing only the realization of the PIOP identity check, we can compare the system-level cost of NTT-based and SumCheck-based approaches without conflating the result with unrelated protocol components such as commitment choices or transcript compression.

\subsection{Univariate ZeroCheck}

Let $\HH \subset \F$ be a multiplicative subgroup of a finite field $\F$, typically of size $N = 2^n$.
Given a polynomial $f(x) \in \F[x]$ (possibly composed of multiple smaller polynomials), the univariate ZeroCheck \cite{gabizon2019plonk} verifies that $f(x)$ vanishes over $\HH$:
$
f(x) = 0 \;\forall x \in \HH 
\;\Leftrightarrow
\exists q(x) \in \F[x] \text{ such that } f(x) = q(x)\cdot z_{\HH}(x),
$
where $z_{\HH}(x)$ is the vanishing polynomial over $\HH$.
When $\HH$ is generated by a root of unity, $z_{\HH}(x) = x^N - 1$.
Thus, the goal of the protocol is to certify that $f(x)$ is divisible by $z_{\HH}(x)$.

The prover computes the quotient $q(x) = f(x)/z_{\HH}(x)$ and commits to it.
A random challenge $r$ is derived by hashing the commitments to $f(x)$ and $q(x)$, after which the prover opens $y=f(r)$ and $t=q(r)$.
The verifier checks whether
$
y = z_{\HH}(r)t,
$
which implies $z_{\HH}(x)\mid f(x)$ except with negligible probability.
Indeed, for a polynomial $p(x)$ of degree $d$, $\Pr_{t \xleftarrow{u} \mathbb{F}}[p(t)=0]\le d/|\mathbb{F}|$, and an invalid quotient would be accepted only if $r$ is a root of $f(x)-q(x)z_{\HH}(x)$,
i.e., $\Pr[f(r) = q(r)\cdot z_{\HH}(r) \mid z_{\HH}(x)\ \text{not divides}\ f(x)]\leq \frac{\text{deg}(f)}{|\mathbb{F}|}$.
The quotient degree satisfies $\deg(q)\le \deg(f)-|\HH|$, so computing $q(x)$ via inverse NTT depends directly on $\deg(f)$, with higher-degree polynomials requiring larger NTT sizes.

% One disadvantage of univariate ZeroCheck is the quotient construction.
% In many ZKP workloads, the original witness polynomials are structured or sparse, meaning that many evaluation points are zero.
% However, once the prover forms $\hat q$ through element-wise division, this sparsity is typically destroyed, thus the resulting quotient polynomial $q$ becomes dense.

\subsection{Multilinear ZeroCheck}
\label{sec:multilinear_zc}

Let $f:\{0,1\}^n\rightarrow\mathbb{F}$ be a function over the Boolean hypercube.
There exists a unique multilinear polynomial $\tilde f\in\mathbb{F}[x_1,\ldots,x_n]$, called the \emph{multilinear extension} (MLE) \cite{thaler_proofs_args_zk, hyperplonk}, such that $\tilde f(x)=f(x)$ for all $x\in\{0,1\}^n$.
A polynomial is {multilinear} if \textit{each} variable $x_i$ has degree at most one.
A multilinear polynomial is fully represented by its $2^n$ evaluations over the hypercube, which we refer to as the \emph{MLE table} \cite{zkspeed2025}.
The MLE admits the interpolation form
$
\tilde f(x)=\sum_{b\in\{0,1\}^n} f(b)\cdot \eq_x(b),
$
where $\eq_x(b)=\prod_{i=1}^n\big((1-b_i)(1-x_i)+b_i x_i\big)$.
The construction of $\eq_x(\cdot)$ is referred to as \emph{Build MLE}
\cite{spartan, zkspeed2025}, 
as it enables evaluating the MLE at an
arbitrary point $x \in \mathbb{F}^n$ from the MLE table \cite{rothblum2024note,bagad2025speeding,dao2024more, dao2024constraint}.

The multilinear ZeroCheck verifies that $f_M$ vanishes on the entire hypercube by checking
\begin{equation}
\label{eq:multilinear_zc_sc}
\sum_{x\in\{0,1\}^n} f_M(x)\cdot \eq_r(x)=0
\end{equation}
for a randomly sampled $r\in\mathbb{F}^n$.
This condition is enforced using the SumCheck protocol.
As shown in \autoref{fig:side-by-side-zc}, $f_M$ is a \textit{combination} of multilinear polynomials, not necessarily multilinear for itself.
Each SumCheck round folds the MLE table along one variable, producing a low-degree polynomial $G_i(X_i)$ and a verifier challenge $\alpha_i$.
After $n=\log_2 N$ rounds, the prover opens the witness polynomials at $\alpha=(\alpha_1,\ldots,\alpha_n)$.
The protocol certifies correctness with soundness error at most 
$\frac{n}{|\mathbb{F}|}$.
\autoref{fig:side-by-side-zc} shows overall procedure.

\begin{figure}[t!]
\centerline{\includegraphics[width=1.03\columnwidth]{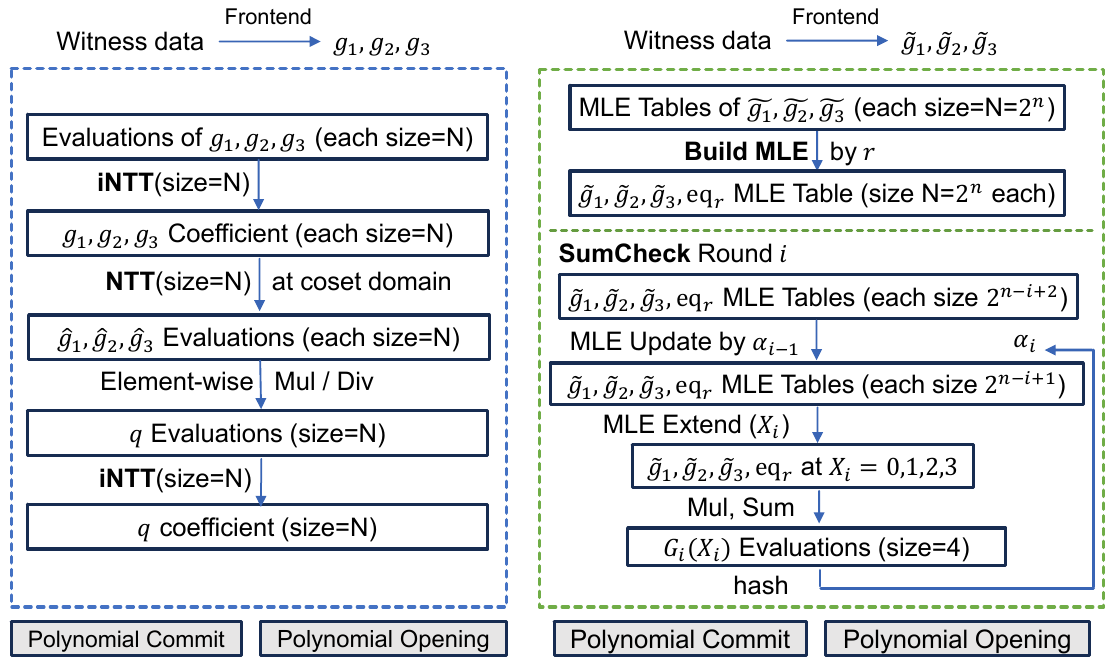}}
\caption{System-level dataflow of ZeroCheck on polynomial $f=g_1 g_2 + g_3$.
Left: univariate ZeroCheck using NTT, polynomial $g_i$ has degree-$N$;
Right: multilinear with SumCheck, $\tilde g_i$ has $n$ variables.
}
\vspace{-1em}
\label{fig:hardware-dataflow}
\end{figure}

\subsection{NTT vs. SumCheck as PIOP}
\label{sec:NTT vs. SumCheck as PIOP}

\autoref{fig:side-by-side-zc} and \ref{fig:hardware-dataflow} illustrate the dataflow of ZeroCheck for the example polynomial $f=g_1 g_2+g_3$, which appears in protocols such as Groth16 \cite{groth16} and Spartan \cite{spartan}.
Both univariate and multilinear ZeroCheck operate on the same number of inputs $N=2^n$.
The key difference lies in representation: the univariate approach encodes each input as a degree-$N$ polynomial, while the multilinear approach represents each as an $n$-variable MLE.
Accordingly, the PIOP layer is realized either by an NTT (univariate) or by a SumCheck core (multilinear).
Here we focus specifically on the PIOP, as both univariate and multilinear realizations commit to the same inputs and thus incur comparable commitment costs.
Moreover, one can adopt lightweight hash-based commitment schemes \cite{ligero_pcs}, which make the additional commitment latency relatively small compared to the overall proving time.

In the univariate setting, correctness is reduced to proving that $z_{\HH}(x)=x^N-1$ divides $f(x)$.
Operationally, the prover interpolates the witness polynomials via INTT, evaluates them over an NTT-friendly coset $D$, and computes a quotient polynomial through element-wise division.
The verifier checks a single random evaluation, yielding a regular $O(N\log N)$ computation with highly regular dataflow.

In the multilinear setting, 
the prover begins with a 
composition of multilinear polynomials over the Boolean hypercube. 
Correctness is reduced to a single check (\autoref{eq:multilinear_zc_sc}) enforced through the SumCheck protocol.
Each SumCheck round, the verifier checks 
$
G_i(0)+G_i(1)=G_{i-1}(r_{i-1})
$ before deriving the next challenge $r_i$.
After $n=\log_2 N$ rounds, the prover opens the witnesses at the derived point.
The computation consists of sequential rounds with lightweight arithmetic and streaming updates to the MLE table.

In both realizations, the prover commits to the underlying witness polynomials (univariate: $g_1,g_2,g_3$; multilinear: $\tilde g_1,\tilde g_2,\tilde g_3$) together with auxiliary quotients or folded polynomials, derives challenges via Fiat-Shamir \cite{fiat1986prove}, and proves consistency through a small number of polynomial openings.
Thus, both ZeroChecks achieve the same logical objective but with distinct computational profiles.

At an algorithmic level, NTT is more compute-intensive than SumCheck, but it yields constant-size proofs, whereas SumCheck produces proofs of size $O(\log N)$.
From a hardware viewpoint, NTT is often considered locality-friendly because data loaded into on-chip memory can be reused across many butterfly stages, while SumCheck repeatedly streams and updates the MLE table across rounds, making it appear to be ``bandwidth-bound''.
However, this intuition is incomplete: the observed performance depends on workload structure (e.g., degree and number of terms) and system constraints (on-chip SRAM capacity and off-chip bandwidth), which can shift the bottleneck and reverse the expected advantage.

\subsection{Performance Drivers and Impact Factors}
\autoref{fig:software-time} shows CPU runtime breakdowns for ZeroCheck with $N=2^{20}$ inputs using NTT- and SumCheck-based PIOPs.
The baseline was executed on an Intel Xeon Gold 5218 \cite{intel_xeon} with 64 threads, a 22 MB cache and 512 GB of main memory.
Although both operate on the same inputs, their costs differ in how polynomial degree and structure translate into computation.

For the example $f=g_1 g_2+g_3$, the univariate case has maximum degree $2N$, whereas the multilinear case $f_M=\tilde g_1\tilde g_2+\tilde g_3$ has constant degree 2 but spans $n$ variables.
In the NTT-based approach, higher degree directly increases transform size:
since $\deg(q)=\deg(f)-\deg(z_\HH)$, letting $\deg(f)=dN$ requires NTTs and iNTTs over $(d-1)N$ points to compute the quotient $q$, in addition to the transforms for each input polynomial ($d=2$ for the above example).
As reflected in the breakdown, input NTTs and $q$ computation dominate the runtime.
In contrast, higher degree SumCheck results in more extensions needed per round, but each extension is computed with lightweight modular additions.
Consequently, increased degree has a stronger impact on NTT computation volume than on SumCheck.

The number of unique polynomials (or additive terms) also impacts performance differently.
With more unique polynomials $g_i$, NTT requires additional transforms for each polynomial, while SumCheck requires more MLE tables to be loaded, updated and processed across rounds.
A further distinction is that univariate ZeroCheck contains element-wise multiplications or divisions.
These operations can be accumulated to prepare $q$'s evaluation and pipelined to save off-chip movement.
In SumCheck, a unique step is Build~MLE $\eq_{r}(x)$ which is computed efficiently with tree-based algorithms and integrated into the SumCheck pipeline \cite{mo2025mtu, zkspeed2025}.

Overall, the CPU results show that NTT incurs longer runtime than SumCheck for this workload.
This is expected on commodity CPUs, where modular arithmetic quickly saturates compute resources, causing the higher computational complexity of NTT to translate directly into longer runtime.
While both approaches incur software overheads, their sources differ: NTT is dominated by large transforms, whereas SumCheck is dominated by Build MLE, MLE table summations, and updates.
These results motivate the hardware evaluation in the following sections, where we examine how these bottlenecks shift under different memory, bandwidth and compute configurations.

%% file: 04Arch.tex
\section{System Design Choice}
\label{sec:architecture}

In this section, we discuss how to design hardware systems for NTT-based PIOPs and SumCheck-based PIOPs. 
The two approaches have different computational structures and dataflow requirements, which lead to distinct system design choices.

\begin{figure}[t!]
\centerline{\includegraphics[width=0.95\columnwidth]{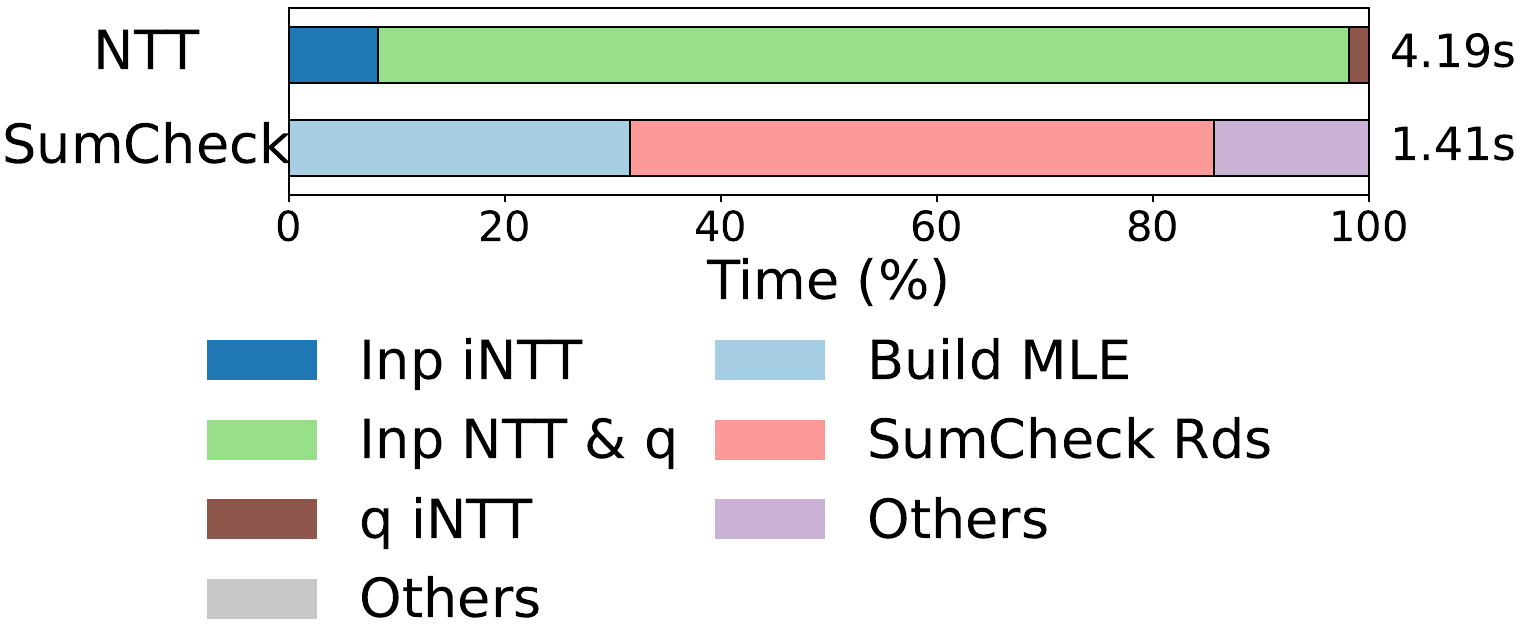}}
\caption{Multilinear (SumCheck) vs. univariate (NTT) CPU runtime breakdown (corresponding to the dash-box in \autoref{fig:hardware-dataflow}). 
Testing workload $f=g_1 g_2+g_3$ with size N=$2^{20}$.
\textit{Inp NTT \& q} includes NTTs and the element-wise calculations.
\textit{Others} in SumCheck contain the runtime evaluating the MLEs at a given round challenge.
}
\vspace{-.3em}
\label{fig:software-time}
\end{figure}

\begin{figure*}[t!]
    \centering
    \begin{minipage}{0.4\textwidth}
        \includegraphics[width=0.96\textwidth]{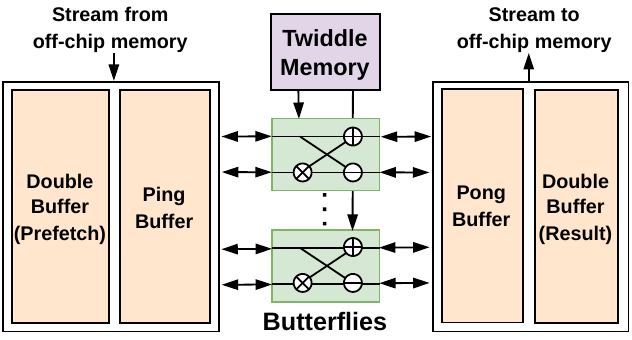}
        \centering
        \caption{~~~~Overview architecture of NTT system}
        \label{fig:ntt_arch}
    \end{minipage}
    \hfill
    \begin{minipage}{0.55\textwidth}
        \centering
        \includegraphics[width=\textwidth]{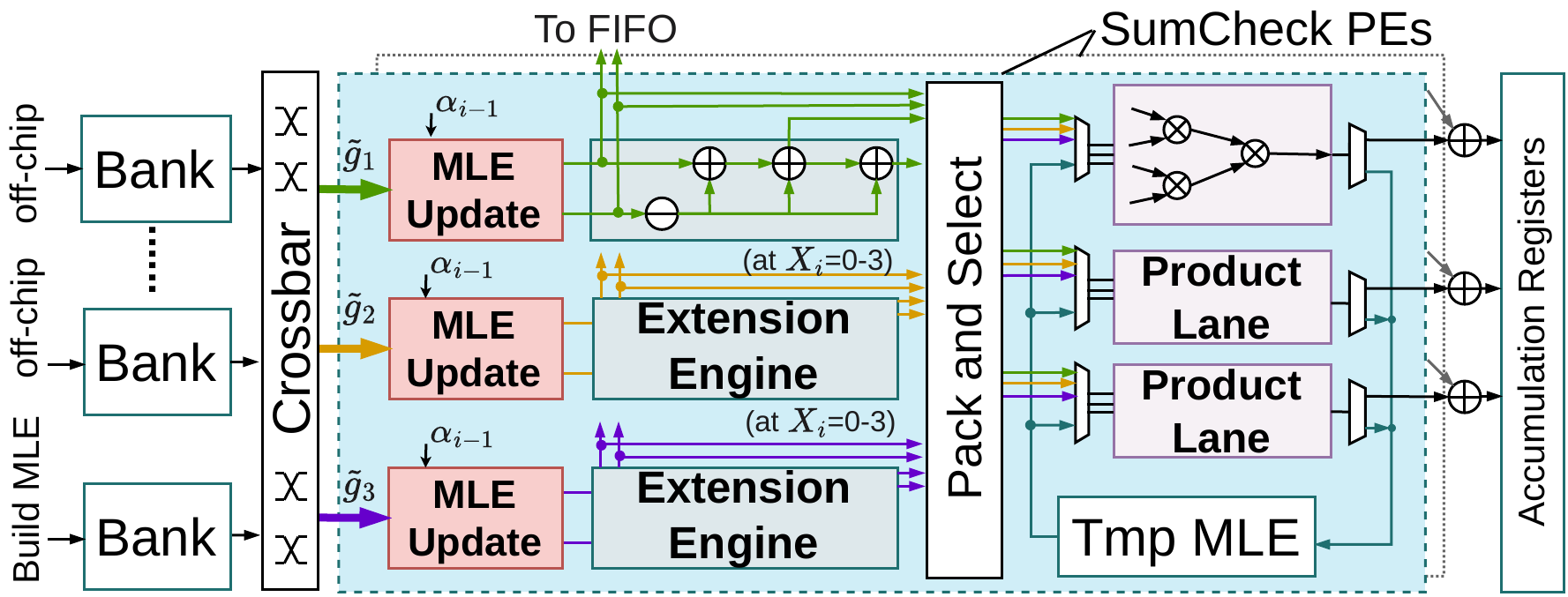}
        \caption{Overview architecture of SumCheck system}
        \label{fig:sumcheck_arch}
    \end{minipage}
\end{figure*}

\subsection{Architecture Choice}

\textbf{NTT}: We use an NTT architecture similar to that discussed in \cite{szkp}. 
This is a {memory}-based architecture in which the initial input is loaded into memory buffers and then \textit{ping-ponged} between buffers across the $\log N$ stages of the transform. 
At each stage, butterfly units read data from one buffer, perform modular multiplications and additions, and write the results to the other buffer. 
We additionally leverage the \textit{constant-geometry} topology \cite{pease1968adaptation} which simplifies the  memory addressing (same data pattern between each stage), ensuring that we avoid additional area overheads from complex scheduling circuits seen in prior works that rely on the Cooley-Tukey NTT \cite{pipezk, legozk, cooley}.
This architecture additionally provisions double buffers to prefetch inputs and transformed outputs so as to overlap computation with memory accesses when possible.

\textbf{SumCheck}: We base the SumCheck architecture on the recently proposed work \cite{zkphire}.
As shown in \autoref{fig:sumcheck_arch}, this architecture similarly reads in polynomials (i.e., MLE tables) into on-chip memories. 
In each round of SumCheck, each polynomial's table entries (i.e., Boolean hypercube instances) are fed simultaneously into the SumCheck pipeline where their extensions and products are computed before being accumulated into registers that store the summation over the Boolean hypercube.
For all rounds except the first round, the polynomials are updated with hash challenges (generated from the summations of the prior round) prior to computing the extensions and products. This effectively reduces the table size of each polynomial to half the size at the beginning of each round.
Additionally, this SumCheck architecture also contains built-in support for construction of $\eq_r(x)$, i.e., the \textit{Build MLE} operation discussed in \autoref{sec:multilinear_zc}.
Construction of $\eq_r(x)$ happens only in the first SumCheck round; subsequent rounds just use the resulting polynomial for updates, extensions, and products.

For both architectures, depending upon the on-chip memory capacity (as we will discuss shortly), the polynomials may be stored entirely on-chip or partially on-chip and partially in off-chip memories. We consider three major factors in the architectural exploration for accelerating the ZeroCheck protocols:
\textit{Compute area} is primarily determined by the number of NTT butterfly units in the NTT design, and by the number of Update and Extension Engines or Product Lanes in the SumCheck design. We additionally consider the number of processing elements (PEs) for both, since allocating multiple PEs is how both architectures support parallel execution where possible. 
\textit{On-chip SRAM} availability controls how much of the working set can be retained locally. 
\textit{Off-chip memory bandwidth} dictates how quickly data can be fetched from or written back to external memory. Together, these resources define the operating regime of the hardware.

\subsection{Three SRAM Scenarios}

\begin{figure}[t!]
\centerline{\includegraphics[width=1\columnwidth]{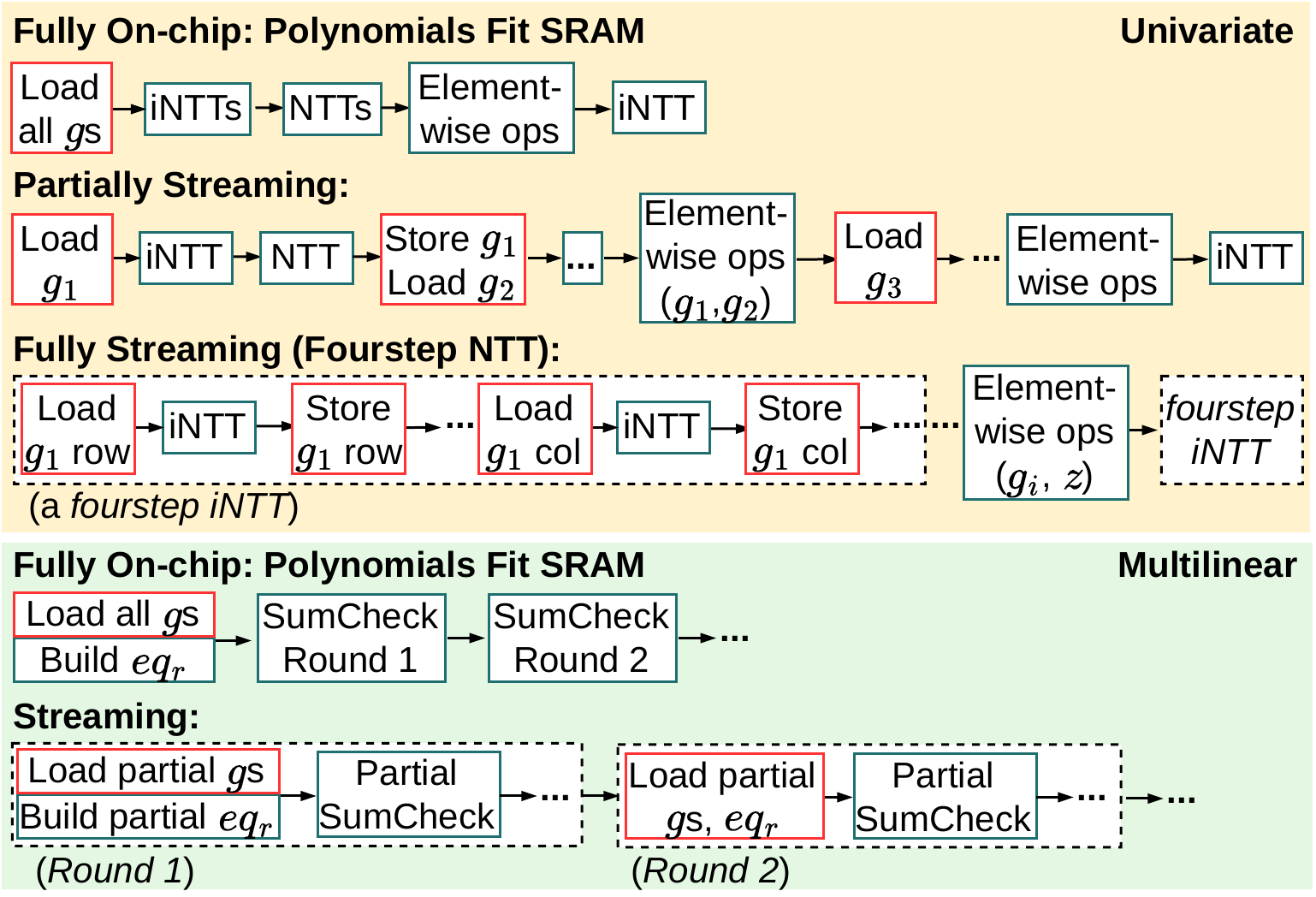}}
\caption{Schedules under the three SRAM scenarios discussed in this section. As workload size grows relative to available SRAM, execution shifts from fully on-chip to streaming with different access patterns.}
\label{fig:scenario-schedule}
\end{figure}

Depending on the available on-chip SRAM and the demands of a given workload, we identify three scenarios for both NTT and SumCheck accelerators, assuming both are provisioned with comparable SRAM capacity, as summarized in \autoref{fig:scenario-schedule}.

\textbf{Scenario I--Fully on-chip.} If sufficient SRAM is available, the entire workload fits on-chip with intermediate data movement to and from off-chip memory. In this case, for a given polynomial, the univariate scheme has all (I)NTTs and related operations run in parallel on-chip, with only the initial inputs and final outputs read/written to off-chip memory. The same is true for the multilinear scheme using SumChecks.
All rounds of SumCheck are performed without intermediate reads and writes from/to off-chip memory.
This scenario is suitable for small workloads, with low-degree polynomials consisting of a limited number of unique polynomials.

\textbf{Scenario II--Partially Streaming.} If SRAM is relatively large but not enough to hold the entire workload, a \emph{streaming mode} is used. The NTTs are computed in a streaming fashion where only one transform can fit fully on-chip. At this SRAM scale, depending upon the input composite polynomial, SumCheck may be able to store all constituent polynomials on-chip, or may need to stream in chunks of polynomials every round (until the reduced polynomials do fit on-chip).

\textbf{Scenario III--Fully Streaming.} When the workload size far exceeds the on-chip SRAM capacity, the system must fully rely on streaming techniques. For NTT, this involves four-step \cite{szkp, pipezk, clake, ark}. 
The Four-Step Bailey's NTT algorithm involves reshaping the original $N$-point input into a $\sqrt N\times\sqrt N$ matrix, and performing column-wise and row-wise NTTs on $\sqrt N$-point
input sequences \cite{fourstep}. In this case, only the $\sqrt N$-point sequence needs to fit on chip (we henceforth refer to these as \textit{mini-NTTs}), at the cost of far more frequent accesses to off-chip memory fetch columns and rows of the matrix into the on-chip buffers, as seen in \autoref{fig:ntt_arch}. Given the frequent memory accesses, we use prefetch and result buffers as in \cite{szkp} for reads and writes to memory. Under sufficient bandwidth constraints, this scheme allows for masking of memory access latencies; at low bandwidth constraints, memory prefetch/writeback latencies may still dominate the total latency.
Four-Step NTTs require additional twiddle factors that are omitted in \autoref{fig:ntt_arch}; we assume additional buffers are provisioned to store and generate these values on-the-fly as done in \cite{szkp}.

\subsection{Degree scaling}
Additional considerations must be made for univariate schemes, where NTTs undergo degree expansion in high-degree polynomials.
In these situations, the resultant quotient degree may not be a power-of-2.
Here, step-radix NTTs \cite{libsnark} are used to decompose a non-power-of-2 length NTT into smaller, power-of-2 sized NTTs, incurring additional element-wise computations.
For example, a polynomial $f = g_1g_2g_3g_4$ might have its constituent univariate $g_i$ polynomials be of degree $N$ (assumed to be a power of $2$), but the final quotient polynomial $q$ is of degree $3N$, which is not a power of $2$. Here, the step radix scheme transforms the $3N$-length sequence into $2N$ and $N$-length sequences, with additional computations to combine the results together to yield a $3N$-length sequence.
Depending upon memory availability, these additional computations may, or may not be sensitive to memory bandwidth

\subsection{Workload scaling and bottlenecks}
Across workload scales, system performance is determined by the balance between compute and memory. Depending on the amount of compute units provisioned and the off-chip bandwidth, the accelerator may become compute-bound or memory-bound.

At smaller workload scales, i.e., when all polynomials fit on-chip, both NTT and SumCheck benefit from additional compute resources, NTT in particular because it has a higher operational intensity (number of arithmetic operations per byte of data fetched from memory) owing to its higher computational complexity.
For very-large workloads, where Scenario III is in play, both NTTs and SumChecks incur a high frequency of memory accesses and are therefore highly sensitive to memory bandwidth.

At workload scales where partial streaming is necessary, the system performance becomes more sensitive to the protocol characteristics along with bandwidth availability.
For univariate based schemes, the computationally heavy operations (NTTs of distinct polynomials) can be performed independently.
This means that even though only a single polynomial of length $N$ can fit on-chip, there is still relatively high per-polynomial data reuse across the $log(N)$ rounds (i.e., a higher operational intensity).
In contrast, for multilinear schemes, the computationally heavy operations are SumCheck rounds over products of multiple $N$-length polynomials, where \textit{all} polynomials in a product must be available simultaneously.
If the on-chip memory can only support $N$ elements, the prover must \textit{tile} the computation, i.e., stream chunks of multiple polynomials in sequence into available on-chip memory.
This tiling preserves correctness and handles the interdependency among polynomials, but lowers locality and makes performance more sensitive to memory bandwidth.

Of course, these observations hold for NTTs and SumChecks of length $N$.
For higher degree polynomials, NTT's undergo degree expansion, in which case the memory capacity to support a single NTT polynomial may be more than enough to support all SumCheck polynomials fully on-chip. 
We examine this high-degree behavior in \autoref{sec:eval}, but even in the simple case of $N$ capacity we already begin to see the complex interplay between memory capacity, bandwidth, and compute allocation in accelerator design.

%% file: 05Evaluation.tex
\section{Evaluation}
\label{sec:eval}

\begin{table}[t]
\centering
\caption{Polynomials evaluated, with their degree, number of unique polynomials, and number of additive terms.
}
\resizebox{0.98\columnwidth}{!}{
\setlength{\tabcolsep}{2mm}{
\rowcolors{2}{rowgray}{white}
\begin{tabular}{lccc}
\toprule
\textbf{Polynomial} & \textbf{Degree ($d$)} & \textbf{Unique Polys} & \textbf{Additive Terms} \\
\midrule
$g_1 g_2$                 & 2 & 2 & 1 \\
$g_1 g_2 + g_3$           & 2 & 3 & 2 \\
$g_1 g_2 + g_3 + g_4$     & 2 & 4 & 3 \\
$g_1 g_2 + g_3 + g_4 + g_5$ & 2 & 5 & 4 \\
$g_1 g_2 g_3$             & 3 & 3 & 1 \\
$g_1 g_2 g_3 g_4$         & 4 & 4 & 1 \\
\bottomrule
\end{tabular}
}
}
\label{tab:poly_examples}
\end{table}

\begin{table}[t]
\centering
\caption{Area and power of primitives}
\label{tab:componentwise_area_power}
\resizebox{0.75\columnwidth}{!}{
\setlength{\tabcolsep}{1mm}{
\begin{tabular}{|c|c|c|}
\hline
\textbf{Component} & \textbf{Area in 7nm} & \textbf{Power} \\ \hline \hline
\textbf{Modular Adder}         & 555 $\mu$m$^2$ & 0.49 mW \\ \hline
\textbf{Modular Multiplier}    & 0.073 mm$^2$ & 63.58 mW \\ \hline
\textbf{1 MB SRAM}             & 0.489 mm$^2$  & 66.67 mW \\ \hline
\end{tabular}
}
}
\vspace{-0.03em}
\end{table}

\subsection{Methodology}
\label{sec:Methodology}

Our evaluation begins with a software baseline on a commodity CPU platform.
We run experiments on an Intel Xeon Gold 5218 \cite{intel_xeon} with 64 threads, an approximate die size of 364\,mm$^2$, and 512\,GB main memory.
For cryptographic primitives we rely on the arkworks library \cite{arkworks}, which provides modular arithmetic and polynomial routines widely used in ZKP systems.

To model hardware execution, we developed a simulator that supports both analytical modeling and cycle-accurate evaluation of NTT and SumCheck PIOPs.
Since both primitives follow fixed, data-oblivious dataflows, their performance can be expressed analytically and validated against detailed pipeline models.
We sweep parameters across workloads and hardware settings.
For workloads, we vary the number of polynomial terms, the maximum polynomial degree, the number of distinct polynomials, and the total input size.
Polynomials are listed in \autoref{tab:poly_examples}, which covers increasing the degree (multiplicative terms) and the number of additive terms. 
We select the polynomials to sweep their properties, while also trying to cover the real-world case. 
For example, $f=g_1g_2+g_3$ has Groth16 \cite{groth16} and Spartan \cite{spartan} style; 
$f=g_1g_2+g_3+g_4+g_5$ is similar to HyperPlonk OpenCheck \cite{hyperplonk};

We test the input workload size from $N=2^{17}$ to $N=2^{32}$.
For hardware, we sweep the available on-chip SRAM capacity and off-chip memory bandwidth, keeping both equal across NTT and SumCheck for a fair comparison.
Within each design, we further explore compute parallelism: in NTT, the number of butterfly units and parallel processing elements (PEs); in SumCheck, the number of MLE Update, Extension Engines, Product Lanes, and Parallel PEs.
Off-chip bandwidth is varied from DDR4-class levels to HBM-class levels to capture a wide performance spectrum.
We assume that both NTT and SumCheck accelerators are integrated into an SoC with a shared DRAM PHY.
To capture real-world scenarios, we test the problem sizes ($N \geq 2^{17}$) that approximate practical ZKP workloads such as Zcash, ZKP-based auctions, and private transaction protocols \cite{zkphire}.

For hardware technology, we prototype compute engines and memory structures using commercial toolchains \cite{haac,mo2023accelerating}. 
We use Catapult HLS 2024 to generate the RTL for Montgomery multipliers, constant-geometry NTT PEs and the fully pipelined SumCheck PE.
The hardware assumes the 255-bit fixed primes used in \cite{zkphire}.
The RTL is synthesized with Synopsys Design Compiler against a TSMC 22\,nm technology library, and SRAM capacity is estimated using a Synopsys 22\,nm Memory Compiler.
The results are scaled to 7\,nm advanced process nodes using standard scaling factors of $3.6\times$ for area and $3.3\times$ for power \cite{szkp, zkspeed2025}, with a target frequency of 1\,GHz.

\subsection{Impact of On-Chip SRAM Availability}
\label{sec:case-Analysis}

We start our analysis of NTT and SumCheck by categorizing workloads based on how much of the computation can fit into on-chip SRAM. We study three representative cases:
(1) {small workloads}, where sufficient SRAM allows the entire workload to be kept on-chip and executed in parallel; 
(2) {medium workloads}, where SRAM can only hold one polynomial at a time to compute its NTT, and execution proceeds in a streaming fashion; and 
(3) {large workloads}, where even a full NTT does not fit and must be decomposed into smaller steps.
SumCheck exhibits similar behavior: with ample SRAM, it runs with all MLEs entirely on-chip (\textit{Full on-chip} mode), taking minimal data movement, while with more limited SRAM it operates in a \textit{streaming} mode, repeatedly loading MLEs from off-chip memory, performing updates, and writing results back.

Across these three cases, we evaluate a range of workloads and highlight the performance trade-offs between NTT- and SumCheck-based hardware, showing how system behavior transitions under different hardware constraints.
To ensure fair comparison, we filter results to the Pareto front, where each design point represents a balance between hardware cost (area) and performance (runtime).
By focusing on Pareto-optimal points for both SumCheck and NTT across all potential designs, we capture systems operating at high resource utilization, avoiding under-provisioned or inefficient configurations.

\subsubsection{Scenario I--Small Workloads}
\label{sec:Fully On-chip Execution}

\begin{figure}[t!]
\centerline{\includegraphics[width=0.93\columnwidth]{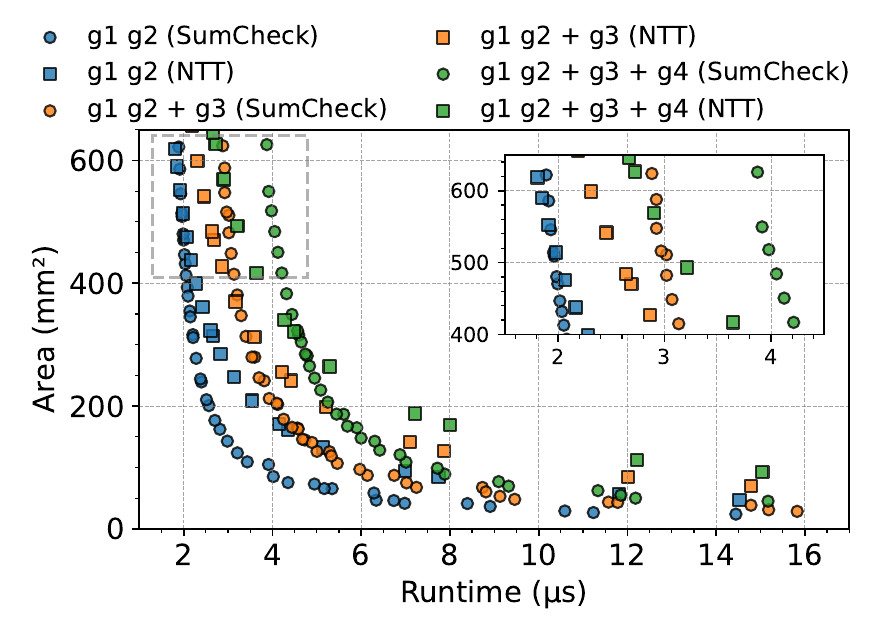}
}
\caption{Pareto Frontiers for small workloads scenario, with increasing number of unique polynomial terms.
At large design points, NTT has better performance than SumCheck.
}
\label{fig:full-on-chip-pareto-3d-more-terms}
\end{figure}

\begin{figure}[t!]
\centerline{\includegraphics[width=1\columnwidth]{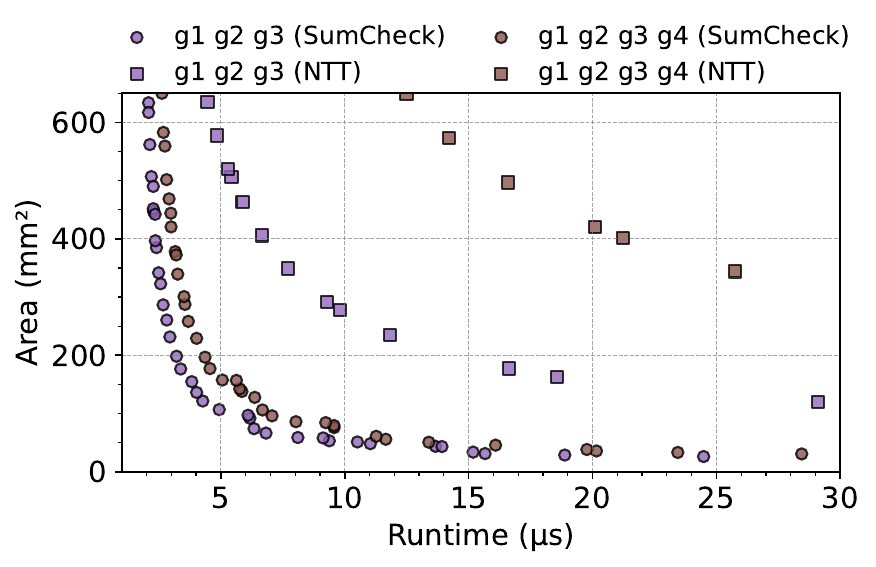}
}
\caption{Pareto Frontiers for small workloads scenario, with increasing polynomial degrees.
As the degree grows, the gap between SumCheck and NTT becomes larger.
}
\label{fig:full-on-chip-pareto-3d-high-degree}
\end{figure}

\begin{figure*}[t!]
\centerline{\includegraphics[width=2.1\columnwidth]{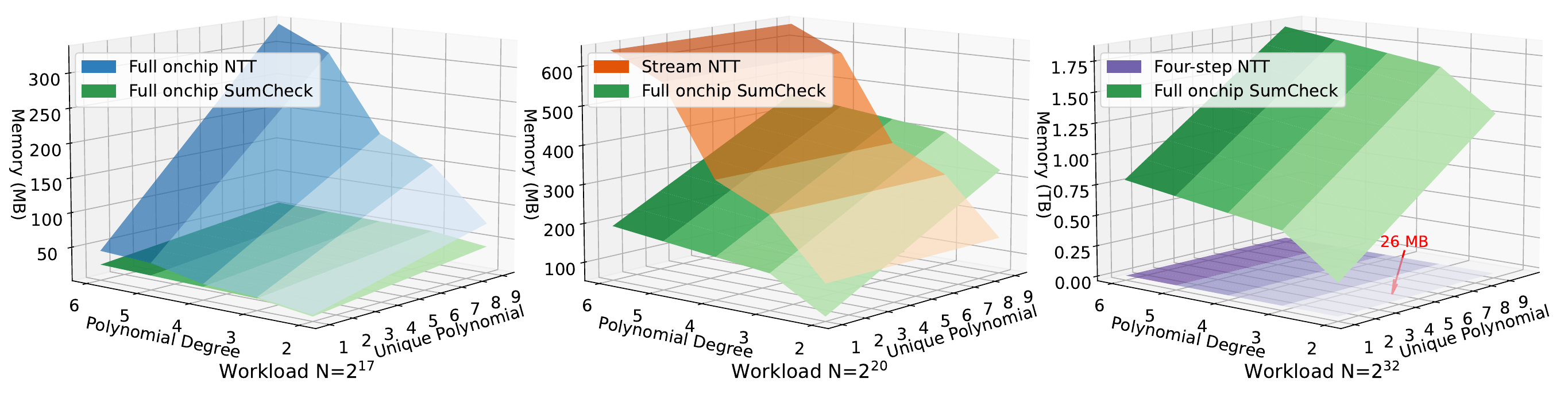}}
\caption{Memory requirement in different scenarios. If the available memory does not reach the requirement (surface in the plot), the system cannot support the corresponding execution.
(a) To fit all polynomials of a small workload full on-chip.
(b) Consider partially streaming NTT. (c) Consider fully streaming (four-step) NTT. The memory requirement for full on-chip SumCheck at this large workload is theoretical. 
}
\label{fig:3d-sram}
\end{figure*}

We begin with the case where the workload is small enough to fit entirely on-chip.
\autoref{fig:full-on-chip-pareto-3d-more-terms} and \autoref{fig:full-on-chip-pareto-3d-high-degree} show the Pareto space analysis for two groups of polynomial workloads: one where the number of additive terms increases, and one where the polynomial degree (i.e., multiplicative terms) increases.
For example, for a workload $f = g_1 g_2$ at size $N = 2^{17}$, an on-chip SRAM budget of 18\,MB is sufficient to store both $g_1$ and $g_2$ entirely on-chip and to execute their NTTs in parallel, without incurring off-chip load and write-back traffic.
Similarly, SumCheck can perform each round within the SRAM buffers, with updated multilinear extensions (MLEs) written back into the same buffer.
This requirement is illustrated in \autoref{fig:3d-sram} (a), which plots the SRAM needed for representative polynomials.
In this example, 18\,MB is enough to support both on-chip NTT and SumCheck.

\textbf{Performance comparison.}
In \autoref{fig:full-on-chip-pareto-3d-more-terms}, we observe that NTT generally exhibits higher latency than SumCheck when the hardware area is small.
This trend follows from the asymptotic compute complexities: NTT requires $O(N \log N)$ operations while SumCheck requires $O(N)$.
With limited compute resources, NTT must perform substantially more arithmetic and therefore runs longer.
At the other extreme, with a large hardware area (zoomed in within \autoref{fig:full-on-chip-pareto-3d-more-terms}), NTT achieves higher peak performance than SumCheck, and the gap becomes more pronounced as the number of polynomial terms increases.
The reason is that in this scenario, with sufficient compute resources, the NTTs' (which operate in parallel) round pipeline is dominated by the latency of parallel butterflies (each butterfly requiring one modular multiplication), while SumCheck rounds 
require not only MLE updates (i.e. two parallel modular multiplications) from the prior challenge, but also further products of extensions. 
This net overhead slows SumCheck relative to NTT.
On the other hand, \autoref{fig:full-on-chip-pareto-3d-high-degree} highlights that for high-degree polynomials with many multiplicative terms, NTT becomes slower.
This is because NTT size scales with polynomial degree.
For example, for $f = g_1 g_2 g_3$ where each $g_i$ is degree-$N$, the quotient polynomial $q$ has degree $2N$, so the NTTs required for each $g_i$ and the INTT required for $q$ must be of size $2N$ rather than $N$.
SumCheck, by contrast, is less sensitive to polynomial degree: increasing degree only requires more evaluation points to represent the polynomial, and the extra evaluations can be computed efficiently using modular additions.
Thus, while higher degrees inflate the NTT size and latency, the impact on SumCheck is modest.

\textbf{On-chip memory requirement.}
\autoref{fig:3d-sram} (a) illustrates the SRAM capacity needed to support full on-chip execution.
If the SRAM budget lies above the NTT or SumCheck surface, then the hardware has sufficient capacity to hold all data structures; otherwise, the workload cannot fit on-chip.
We observe that under this assumption, NTT generally requires more SRAM than SumCheck, especially as the number of unique polynomials and the polynomial degree increase.

Increasing the number of unique polynomials requires additional buffers to store each new polynomial, and NTT incurs further overhead from the ping-pong buffering needed to support its pipeline.
Likewise, higher polynomial degree increases both the NTT size and the buffer space required to store intermediate values.
For example, for $f = g_1 g_2$ the system must compute an $N$-size NTT for the quotient $q$, while for $f = g_1 g_2 g_3$ the required NTT size doubles to $2N$ due to the increased degree.
In contrast, SumCheck is less sensitive to polynomial degree: growing the degree does not expand the MLE buffer, but only increases the size of the temporary buffer used to store additional hypercube extension points.
Thus, if the hardware budget is sufficient to accommodate a full NTT on-chip, it can also accommodate SumCheck on-chip.

Due to the use of step-radix NTT, the SRAM requirement does not grow linearly with polynomial degree.
For instance, for $f = g_1 g_2 g_3 g_4 g_5 g_6$ the system needs to compute a $5N$-size NTT, which can be realized as a composition of $4N$- and $N$-size NTTs.
In this case, the SRAM requirement is bounded by the $4N$ instance rather than the full $5N$, reducing the effective memory footprint.

\begin{takeawaybox}
With full on-chip execution, NTT is compute-heavier and slower at small hardware budgets but NTT outperforms SumCheck at larger scale, due to the shorter round pipeline.
NTT's latency and SRAM demand grow sharply with polynomial degree.
By contrast, SumCheck scales more gently, so any SRAM capacity sufficient for an NTT will be sufficient for SumCheck.
\end{takeawaybox}

\subsubsection{Scenario II--Medium Workloads}
\label{sec:medium-workload}

\begin{figure*}[t]
\centerline{\includegraphics[width=2\columnwidth]{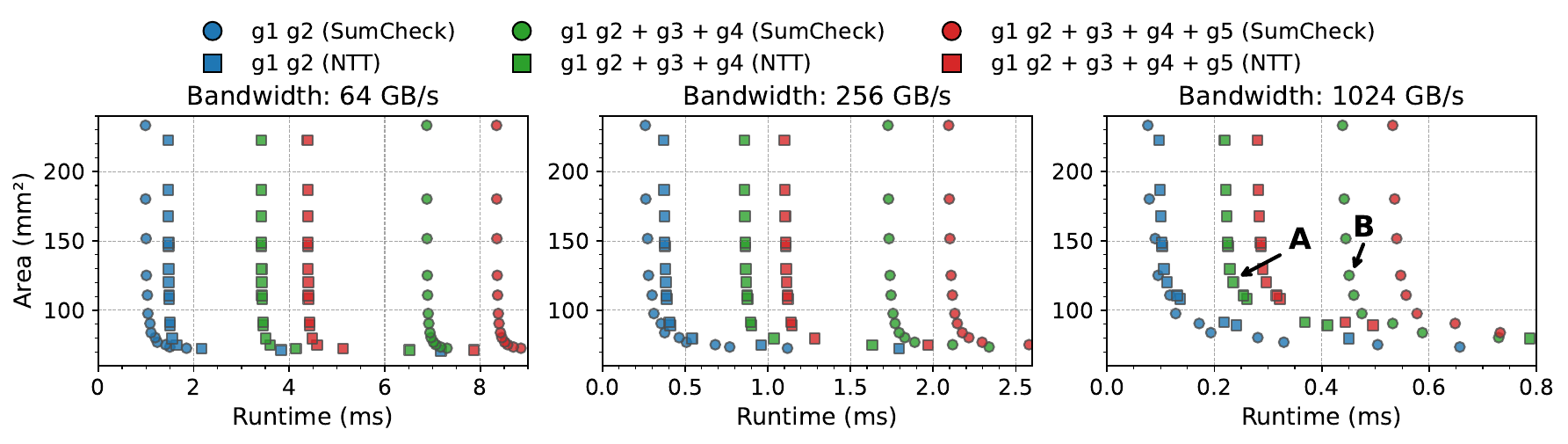}}
\caption{Pareto plots of NTT and SumCheck architectures under Scenario II, where we examine the effect of increasing the number of unique polynomials while keeping the degree fixed. Workload size $N=2^{20}$.
}
\label{fig:streaming-pareto1}
\end{figure*}

\begin{figure*}[t]
\centerline{\includegraphics[width=2\columnwidth]{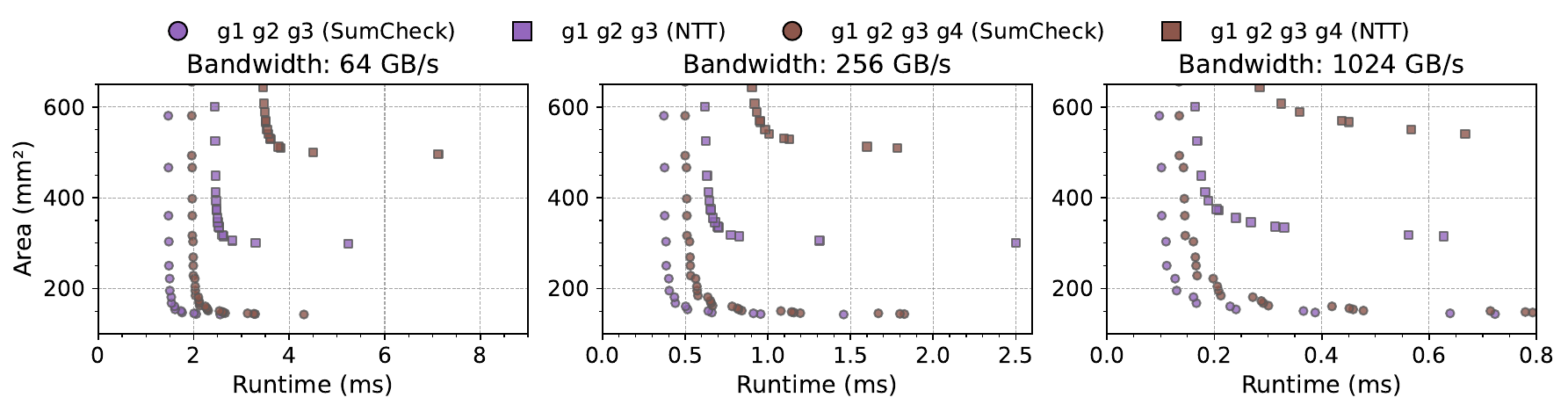}}
\caption{Pareto plots of NTT and SumCheck architectures under Scenario II, where we examine the effect of increasing the total polynomial degree. Workload size $N=2^{20}$.
}
\label{fig:streaming-pareto2}
\end{figure*}

We next consider a medium-size workload with $N=2^{20}$.
In this setting, the on-chip memory is insufficient to hold all polynomials simultaneously, but it can still accommodate one polynomial and its NTT at a time.
This enables a scheduling strategy where NTTs for different polynomials are computed sequentially.
While one NTT is being processed on-chip, the next polynomial is prefetched into a double buffer, overlapping computation and data movement to hide loading latency.

For SumCheck, given the same SRAM budget as NTT, there are two possibilities.
As illustrated in \autoref{fig:3d-sram} (b), when the polynomial degree is high, the SRAM budget required for streaming NTT is larger than the SRAM needed for a full on-chip SumCheck.
In this case, the SumCheck MLEs can all be loaded and stored on-chip, achieving full on-chip execution.
This occurs because higher degree polynomials increase the NTT size and the associated ping-pong buffers, while the SumCheck buffers grow more slowly.
In contrast, when the workload consists of many unique polynomials, the SRAM budget sufficient for streaming NTT may still fall short of fitting all polynomials required for full on-chip SumCheck.
For example, with $f = g_1 g_2 + g_3 + g_4$, four unique polynomials are required.
A streaming NTT only needs buffers to support one NTT at a time, while SumCheck requires simultaneous access to all polynomials during each round to update and generate new MLEs.
As a result, full on-chip execution is not possible for SumCheck in this case, and the design must operate in a streaming mode that stores partial chunks of each polynomial on-chip.

This understanding helps interpret the Pareto analysis results in \autoref{fig:streaming-pareto1}, which evaluates workloads with increasing numbers of unique polynomials.
For workloads with few polynomials (e.g., $f = g_1 g_2$), SumCheck outperforms NTT because the given SRAM capacity is already sufficient to fit all polynomials on-chip, while NTT remains bandwidth-limited.
As bandwidth increases, the performance gap between SumCheck and NTT narrows.
However, as the number of unique polynomials grows (e.g., $f = g_1 g_2 + g_3 + g_4$), both NTT and SumCheck must operate in streaming mode.
In this regime, NTT achieves better performance at the same bandwidth, reflecting its stronger data reuse.
Streaming NTT can load a polynomial on-chip and complete \textit{all} its rounds while prefetching the next polynomial, whereas streaming SumCheck must load \textit{partial} chunks of all polynomials, compute a single round across them, and then write intermediate results back off-chip.
Unlike NTT, SumCheck cannot begin the next round until all MLEs are processed at current round. 
In other words, SumCheck cannot load one MLE and compute all its rounds, or it cannot load the partial across all MLEs to compute all their rounds. 
This repeated on/off-chip movement lowers performance.

\autoref{fig:streaming-pareto2} further illustrates workloads with increasing multiplicative terms (higher degree).
As the NTT size grows with degree, the SRAM budget allocated for streaming NTT already suffices to support SumCheck fully on-chip.
In this scenario, SumCheck achieves better performance.
We also observe that NTT exhibits a higher minimum area cost at low-performance points, since the required on-chip SRAM for large NTT sizes includes ping-pong buffers, twiddle factors, and staging buffers that collectively exceed the buffer costs for SumCheck MLEs and temporary storage.

\begin{takeawaybox}
SumCheck outperforms NTTs for high-degree polynomials due to the degree expansion experienced by NTT protocols, meaning that on-chip capacity that can support one NTT on-chip can support the corresponding SumCheck fully on-chip. However, interestingly, at lower degrees with more unique polynomials, NTTs outperform SumChecks because of SumCheck's per-round dependency on \textit{all} polynomials, incurring higher memory traffic, while NTTs can run \textit{independently} in series and exploit greater data locality.
\end{takeawaybox}

\subsubsection{Scenario III--Large Workloads}
\label{sec:large-fourstep}

For large workloads such as $N = 2^{32}$, it is no longer feasible to provide sufficient on-chip SRAM to store either all polynomials or even a single polynomial in its entirety.
\autoref{fig:3d-sram} (c) shows the theoretical SRAM requirement if we attempted to store all polynomials fully on-chip.
To handle such scales, NTT computation must be decomposed using the four-step Bailey NTT algorithm \cite{fourstep}.
In this scheme, a full $N = 2^{32}$ transform is executed by composing many $N = 2^{16}$ mini-NTTs.

\begin{figure*}[t!]
\centerline{\includegraphics[width=2\columnwidth]{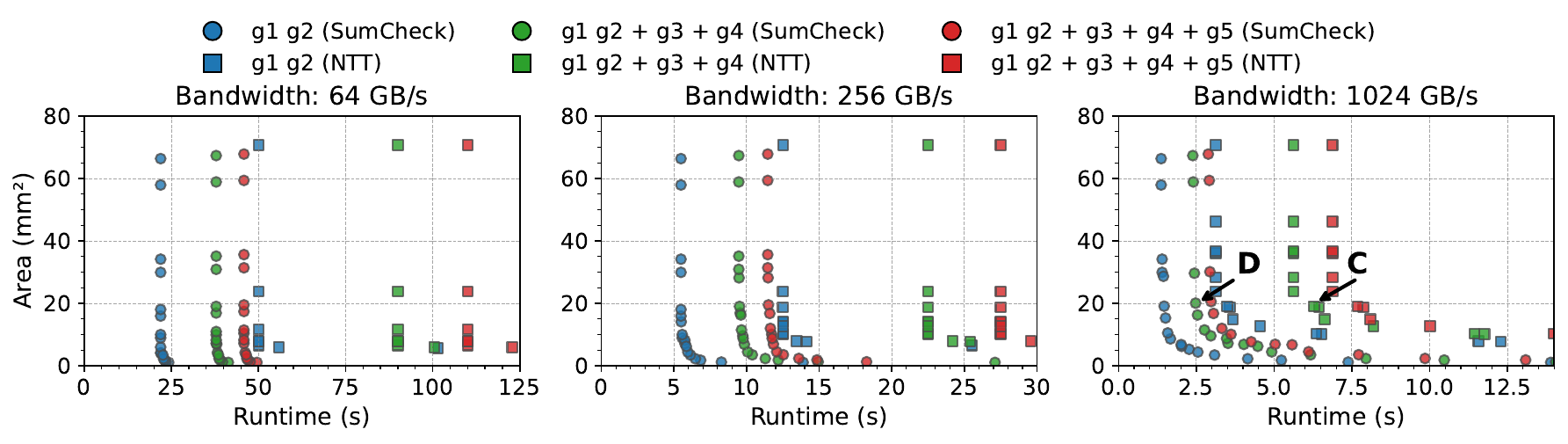}}
\caption{Pareto plots of NTT and SumCheck systems under Scenario III (four-step NTT).
We examine the effect of increasing the number of unique polynomials. Workload size $N=2^{32}$.
}
\label{fig:four-step-pareto}
\end{figure*}

\autoref{fig:four-step-pareto} illustrates that, under this setting, SumCheck consistently outperforms NTT across different polynomial configurations.
Increasing off-chip bandwidth improves performance for both primitives, but unlike the medium workload regime (\autoref{sec:medium-workload}), NTT does not regain superiority.
The reason is that the four-step decomposition disrupts NTT's usual data reuse.
Specifically, each column-wise $2^{16}$ chunk must be loaded to perform a mini-NTT, then written back; afterward, each row-wise $2^{16}$ chunk must be loaded for another mini-NTT and written back.
As a result, the system performs two full passes of reading and writing $N=2^{32}$ elements, effectively doubling the off-chip traffic compared to streaming NTT without decomposition.

In streaming SumCheck, the prover must also read the MLE table from memory and update it to produce the reduced table.
This results in an off-chip movement pattern equivalent to two full passes of reading and writing $N=2^{32}$ elements.
However, since NTT loses its reuse advantage under four-step decomposition, its latency becomes higher than SumCheck in this regime.

Finally, as polynomial degree increases, NTT's performance degrades further because the transform size grows proportionally with degree.
For example, higher-degree products yield larger quotient polynomials, and the required NTT size increases accordingly.
In contrast, SumCheck scales more gracefully in this setting, since degree growth does not directly inflate the size of its MLE table.

\begin{takeawaybox}
For very large workloads, NTT requires four-step decomposition, which breaks its data-reuse advantage and increases off-chip traffic. Streaming SumCheck has similar bandwidth costs but scales more gracefully with degree, so it consistently outperforms NTT in this regime.
\end{takeawaybox}

\subsubsection{Performance breakdown}
Finally, we analyze the Pareto front performance points for both medium and large workloads.
\autoref{tab:detail-breakdown} summarizes the breakdown of key components, with four representative samples taken from \autoref{fig:streaming-pareto1} and \autoref{fig:four-step-pareto}.

One key observation is that transitioning from streaming NTT to four-step NTT (points A to C) reduces the amount of compute hardware that can be effectively allocated.
In streaming NTT, data remains on-chip until a full NTT and iNTT are completed, enabling significant reuse and justifying the allocation of more butterfly units.
By contrast, in the four-step decomposition, data chunks must be written back to off-chip memory after each mini-NTT.
As a result, adding more butterflies does not improve performance on the Pareto front, since computation stalls waiting for off-chip transfers.

We also note that the latency of the input (i)NTTs is not directly proportional to the latency of the quotient $q$ iNTT.
This occurs because input transforms include both computation and data movement, whereas $q$’s iNTT can begin directly from element-wise result buffers, avoiding additional transfer overhead.

For medium workloads, the SumCheck system (point B) shows similar cost components to streaming NTT.
However, its runtime remains bounded by off-chip streaming bandwidth.
Even with sufficient compute units, SumCheck cannot exceed the performance ceiling imposed by memory traffic, whereas NTT performance is constrained by the balance between reuse and bandwidth depending on the decomposition strategy.

\begin{table}[t]
\centering
\caption{Detail breakdown of points from \autoref{fig:streaming-pareto1} and \ref{fig:four-step-pareto}, including the design points' area, peak power and runtime (latency).
}
\rowcolors{2}{rowgray}{white}
\resizebox{1\columnwidth}{!}{
\setlength{\tabcolsep}{1mm}{
\begin{tabular}{lrrlr}
\hline
\multicolumn{5}{c}{\textbf{A (NTT)}}                                                                                                                      \\
                     & \multicolumn{1}{l}{\textbf{Area (mm$^2$)}} & \textbf{Power (W)}                     &                      & \textbf{Latency (\%)} \\ \hline
\textbf{Butterflies} & 50.03                                      & \multicolumn{1}{r|}{42.73}             & \textbf{Input (i)NTTs} & 91.9                  \\
\textbf{Ping-pong}   & 31.17                                      & \multicolumn{1}{r|}{4.27}              & \textbf{q iNTT}      & 6.8                   \\
\textbf{Other Bufs}  & 38.96                                      & \multicolumn{1}{r|}{5.33}              & \textbf{Others}      & 1.3                   \\ \hline
\multicolumn{5}{c}{\textbf{B (SumCheck)}}                                                                                                                 \\
                     & \multicolumn{1}{l}{\textbf{Area (mm$^2$)}} & \multicolumn{1}{l}{\textbf{Power (W)}} &                      & \textbf{Latency (\%)} \\ \hline
\textbf{Compute}     & 53.97                                      & \multicolumn{1}{r|}{46.80}             & \textbf{Round 1}        & 28.0                  \\
\textbf{MLE Banks}   & 70.12                                      & \multicolumn{1}{r|}{9.56}              & \textbf{Round 1-2}      & 51.4                  \\
\textbf{Temp Buf}     & 0                                          & \multicolumn{1}{r|}{0}                 & \textbf{Round 3-20}     & 20.6                  \\ \hline \hline
\multicolumn{5}{c}{\textbf{C (NTT)}}                                                                                                                      \\
                     & \multicolumn{1}{l}{\textbf{Area (mm$^2$)}} & \multicolumn{1}{l}{\textbf{Power (W)}} &                      & \textbf{Latency (\%)} \\ \hline
\textbf{Butterflies} & 2.38                                       & \multicolumn{1}{r|}{2.03}              & \textbf{Input (i)NTTs} & 86.8                  \\
\textbf{Ping-pong}   & 11.69                                      & \multicolumn{1}{r|}{1.60}              & \textbf{q iNTT}      & 10.8                  \\
\textbf{Other Bufs}  & 5.36                                       & \multicolumn{1}{r|}{0.73}              & \textbf{Others}      & 2.4                   \\ \hline
\multicolumn{5}{c}{\textbf{D (SumCheck)}}                                                                                                                 \\
                     & \multicolumn{1}{l}{\textbf{Area (mm$^2$)}} & \multicolumn{1}{l}{\textbf{Power (W)}} &                      & \textbf{Latency (\%)} \\ \hline
\textbf{Compute}     & 16.43                                      & \multicolumn{1}{r|}{14.24}             & \textbf{Round 1}        & 24.6                  \\
\textbf{MLE Banks}   & 2.43                                       & \multicolumn{1}{r|}{0.33}              & \textbf{Round 1-2}      & 37.7                  \\
\textbf{Temp Buff}    & 0.49                                       & \multicolumn{1}{r|}{0.07}              & \textbf{Round 3-26}     & 37.6                  \\ \hline
\end{tabular}
}
}
\label{tab:detail-breakdown}
\end{table}

%% file: 06Related.tex
\section{Related Work}
\label{sec:related}

Several prior works \cite{sam, pipezk, gzkp, unizk, legozk, zkpog, szkp} have studied NTT acceleration with ZKPs as a primary use case, alongside a broader body of work on NTT/FFT acceleration in other domains \cite{moma_cgo, moma_micro, rpu, zhu2026efficient}, especially fully homomorphic encryption \cite{ciflow, kim2025anaheim, kim2022_BTS, osiris}. 
These works use a variety of hardware topologies, including constant-geometry and Cooley-Tukey variants, as well as fully-pipelined and memory-based architectures.
This work uses constant-geometry, memory-based NTTs to enable simple designs and eliminate the overhead of complex scheduling logic area inherent to Cooley-Tukey and pipelined architectures.
However, prior work has not explicitly analyzed the hardware-system-level tradeoffs introduced by high-degree polynomials. 
This work has sought to fill in this knowledge gap.

In comparison, fewer works have looked at SumCheck acceleration \cite{zkspeed2025, nocap, zkphire} in hardware.
zkSpeed \cite{zkspeed2025} proposes a custom SumCheck accelerator to handle a few select polynomials tuned to the HyperPlonk \cite{hyperplonk} protocol, while NoCap \cite{nocap} presents a vector processor that accelerates polynomials in the Spartan \cite{spartan} protocol. Both schemes suffice for their use cases but struggle to generalize; zkSpeed cannot support custom, high-degree gates, and reduction operations like folding (i.e., summing a vector) are known to be suboptimal when implemented using vector-based primitives. This work uses the  programmable SumCheck in \cite{zkphire} as it supports
custom high-degree gates with high-degrees of parallelism without incurring the slowdowns for vector-based reductions.

\section{Conclusion}
\label{sec:conclusion}

The ZKP community has discussed the asymptotic costs of SumCheck and NTT: SumCheck offers $O(N)$ complexity while NTT requires $O(N \log N)$.
In parallel, hardware designers have argued that NTT is more accelerator-friendly due to data locality and reuse, whereas SumCheck's sequential, round-based structure stresses memory bandwidth.
Despite these opposing narratives, the two approaches have not been directly compared at the hardware-system level under a unified architectural framework.

In this paper we provide such a comparison, implementing optimized designs for both NTT and SumCheck and evaluating them under the same level SRAM and off-chip bandwidth budgets.
Our analysis shows that while there is no clear winner for \textit{every} situation, SumChecks are generally preferable for larger workload scales and higher-degree polynomials. 
For lower-degree polynomials, NTTs might be more advantageous under certain hardware, workload, and memory configurations. 
These findings bridge protocol design and hardware architecture, offering a different view to protocol builders on selecting NTT or SumCheck in constructing efficient zero-knowledge proof systems.

%% file: ref-sample-base.bib
@String{JACM = "J. ACM" }

@String{Computing = "Computing" }

@String{Computer = "{IEEE} Computer" }

@String{Springer = "Springer-Verlag" }

@inproceedings{mo2025mtu,
  title={MTU: The Multifunction Tree Unit for Accelerating Zero-Knowledge Proofs},
  author={Mo, Jianqiao and Daftardar, Alhad and Ah-Kiow, Joey and Guo, Kaiyue and B{\"u}nz, Benedikt and Garg, Siddharth and Reagen, Brandon},
  booktitle={Proceedings of the 14th International Workshop on Hardware and Architectural Support for Security and Privacy},
  pages={19--27},
  year={2025}
}

@inproceedings{zkspeed2025,
author = {Daftardar, Alhad and Mo, Jianqiao and Ah-kiow, Joey and B\"{u}nz, Benedikt and Karri, Ramesh and Garg, Siddharth and Reagen, Brandon},
title = {Need for zkSpeed: Accelerating HyperPlonk for Zero-Knowledge Proofs},
year = {2025},
isbn = {9798400712616},
publisher = {Association for Computing Machinery},
address = {New York, NY, USA},
url = {https://doi.org/10.1145/3695053.3731021},
doi = {10.1145/3695053.3731021},
booktitle = {Proceedings of the 52nd Annual International Symposium on Computer Architecture},
pages = {1986–2001},
numpages = {16},
location = {
},
series = {ISCA '25}
}

@INPROCEEDINGS{fourstep,
  author={Bailey, D. H.},
  booktitle={Supercomputing '89:Proceedings of the 1989 ACM/IEEE Conference on Supercomputing}, 
  title={FFTs in external or hierarchical memory}, 
  year={1989},
  volume={},
  number={},
  pages={234-242},
  doi={10.1145/76263.76288}}

@inproceedings{haac,
  title={Haac: A hardware-software co-design to accelerate garbled circuits},
  author={Mo, Jianqiao and Gopinath, Jayanth and Reagen, Brandon},
  booktitle={Proceedings of the 50th Annual International Symposium on Computer Architecture},
  pages={1--13},
  year={2023}
}

@inproceedings{mo2023accelerating,
  title={Accelerating Garbled Circuits by Hardware-Software Co-Design},
  author={Mo, Jianqiao and Reagen, Brandon},
  booktitle={DISCC 2023 2nd Workshop on Data Integrity and Secure Cloud Computing},
  year={2023}
}

@software{arkworks,
  author = {arkworks},
  title = {\texttt{arkworks} zkSNARK ecosystem},
  url = {https://arkworks.rs},
  year = {2022},
}

@article{osiris,
  title={Osiris: A systolic approach to accelerating fully homomorphic encryption},
  author={Ebel, Austin and Reagen, Brandon},
  journal={ACM Transactions on Architecture and Code Optimization},
  volume={23},
  number={1},
  pages={1--27},
  year={2026},
  publisher={ACM New York, NY}
}

@inproceedings{clake,
  title={Craterlake: a hardware accelerator for efficient unbounded computation on encrypted data},
  author={Samardzic, Nikola and Feldmann, Axel and Krastev, Aleksandar and Manohar, Nathan and Genise, Nicholas and Devadas, Srinivas and Eldefrawy, Karim and Peikert, Chris and Sanchez, Daniel},
  booktitle={Proceedings of the 49th Annual International Symposium on Computer Architecture},
  pages={173--187},
  year={2022}
}

@INPROCEEDINGS{ark,
  author={Kim, Jongmin and Lee, Gwangho and Kim, Sangpyo and Sohn, Gina and Rhu, Minsoo and Kim, John and Ahn, Jung Ho},
  booktitle={2022 55th IEEE/ACM International Symposium on Microarchitecture (MICRO)}, 
  title={ARK: Fully Homomorphic Encryption Accelerator with Runtime Data Generation and Inter-Operation Key Reuse}, 
  year={2022},
  volume={},
  number={},
  pages={1237-1254},
  doi={10.1109/MICRO56248.2022.00086}}

@inproceedings{szkp,
  title={SZKP: A scalable accelerator architecture for zero-knowledge proofs},
  author={Daftardar, Alhad and Reagen, Brandon and Garg, Siddharth},
  booktitle={Proceedings of the 2024 International Conference on Parallel Architectures and Compilation Techniques},
  pages={271--283},
  year={2024}
}

@inproceedings{kim2022_BTS,
  title={Bts: An accelerator for bootstrappable fully homomorphic encryption},
  author={Kim, Sangpyo and Kim, Jongmin and Kim, Michael Jaemin and Jung, Wonkyung and Kim, John and Rhu, Minsoo and Ahn, Jung Ho},
  booktitle={Proceedings of the 49th annual international symposium on computer architecture},
  pages={711--725},
  year={2022}
}

@inproceedings{hyperplonk,
  title={Hyperplonk: Plonk with linear-time prover and high-degree custom gates},
  author={Chen, Binyi and B{\"u}nz, Benedikt and Boneh, Dan and Zhang, Zhenfei},
  booktitle={Annual International Conference on the Theory and Applications of Cryptographic Techniques},
  pages={499--530},
  year={2023},
  organization={Springer}
}

@article{zkphire,
  title={zkPHIRE: A Programmable Accelerator for ZKPs over HIgh-degRee, Expressive Gates},
  author={Daftardar, Alhad and Mo, Jianqiao and Ah-kiow, Joey and B{\"u}nz, Benedikt and Garg, Siddharth and Reagen, Brandon},
  journal={arXiv preprint arXiv:2508.16738},
  year={2025}
}

@InProceedings{kzg_pcs,
author="Kate, Aniket
and Zaverucha, Gregory M.
and Goldberg, Ian",
editor="Abe, Masayuki",
title="Constant-Size Commitments to Polynomials and Their Applications",
booktitle="Advances in Cryptology - ASIACRYPT 2010",
year="2010",
publisher="Springer Berlin Heidelberg",
address="Berlin, Heidelberg",
pages="177--194",
isbn="978-3-642-17373-8"
}

@INPROCEEDINGS{sam,
  author={Wang, Cheng and Gao, Mingyu},
  booktitle={2023 IEEE/ACM International Conference on Computer Aided Design (ICCAD)}, 
  title={SAM: A Scalable Accelerator for Number Theoretic Transform Using Multi-Dimensional Decomposition}, 
  year={2023},
  volume={},
  number={},
  pages={1-9},
  doi={10.1109/ICCAD57390.2023.10323744}}

@misc{zkpog,
      author = {Muyang Li and Yueteng Yu and Bangyan Wang and Xiong Fan and Shuwen Deng},
      title = {{ZKPoG}: Accelerating {WitGen}-Incorporated End-to-End Zero-Knowledge Proof on {GPU}},
      howpublished = {Cryptology {ePrint} Archive, Paper 2025/765},
      year = {2025},
      url = {https://eprint.iacr.org/2025/765}
}

@INPROCEEDINGS{nocap,
  author={Samardzic, Nikola and Langowski, Simon and Devadas, Srinivas and Sanchez, Daniel},
  booktitle={2024 57th IEEE/ACM International Symposium on Microarchitecture (MICRO)}, 
  title={Accelerating Zero-Knowledge Proofs Through Hardware-Algorithm Co-Design}, 
  year={2024},
  volume={},
  number={},
  pages={366-379},
  doi={10.1109/MICRO61859.2024.00035}}

@article{pease1968adaptation,
  title={An adaptation of the fast Fourier transform for parallel processing},
  author={Pease, Marshall C},
  journal={Journal of the ACM (JACM)},
  volume={15},
  number={2},
  pages={252--264},
  year={1968},
  publisher={ACM New York, NY, USA}
}

@InProceedings{spartan,
author="Setty, Srinath",
editor="Micciancio, Daniele
and Ristenpart, Thomas",
title="Spartan: Efficient and General-Purpose zkSNARKs Without Trusted Setup",
booktitle="Advances in Cryptology -- CRYPTO 2020",
year="2020",
publisher="Springer International Publishing",
address="Cham",
pages="704--737",
isbn="978-3-030-56877-1"
}

@InProceedings{groth16,
author="Groth, Jens",
editor="Fischlin, Marc
and Coron, Jean-S{\'e}bastien",
title="On the Size of Pairing-Based Non-interactive Arguments",
booktitle="Advances in Cryptology -- EUROCRYPT 2016",
year="2016",
publisher="Springer Berlin Heidelberg",
address="Berlin, Heidelberg",
pages="305--326",
isbn="978-3-662-49896-5"
}

@INPROCEEDINGS{sumcheck,
  author={Lund, C. and Fortnow, L. and Karloff, H. and Nisan, N.},
  booktitle={Proceedings [1990] 31st Annual Symposium on Foundations of Computer Science}, 
  title={Algebraic methods for interactive proof systems}, 
  year={1990},
  volume={},
  number={},
  pages={2-10 vol.1},
  doi={10.1109/FSCS.1990.89518}}

@ARTICLE{cooley,
  author={Cooley, James W. and Lewis, Peter A. W. and Welch, Peter D.},
  journal={IEEE Transactions on Education}, 
  title={The Fast Fourier Transform and Its Applications}, 
  year={1969},
  volume={12},
  number={1},
  pages={27-34},
  doi={10.1109/TE.1969.4320436}}

@inproceedings{gzkp,
author = {Ma, Weiliang and Xiong, Qian and Shi, Xuanhua and Ma, Xiaosong and Jin, Hai and Kuang, Haozhao and Gao, Mingyu and Zhang, Ye and Shen, Haichen and Hu, Weifang},
title = {GZKP: A GPU Accelerated Zero-Knowledge Proof System},
year = {2023},
isbn = {9781450399166},
publisher = {Association for Computing Machinery},
address = {New York, NY, USA},
url = {https://doi.org/10.1145/3575693.3575711},
doi = {10.1145/3575693.3575711},
booktitle = {Proceedings of the 28th ACM International Conference on Architectural Support for Programming Languages and Operating Systems, Volume 2},
pages = {340–353},
numpages = {14},
location = {Vancouver, BC, Canada},
series = {ASPLOS 2023}
}

@inproceedings{pipezk,
  author = "Ye Zhang and Shuo Wang and Xian Zhang and Jiangbin Dong and Xingzhong Mao and Fan Long and Cong Wang and Dong Zhou and Mingyu Gao and Guangyu Sun",
  title = "PipeZK: Accelerating Zero-Knowledge Proof with a Pipelined Architecture",
  year = "2021",
  booktitle = "2021 ACM/IEEE 48th Annual International Symposium on Computer Architecture (ISCA)"
}

@software{libsnark,
  title = "libsnark: a C++ library for zkSNARK proofs.",
  year = "2018",
  url = "https://github.com/scipr-lab/libsnark"
}

@software{libfqfft,
  title = "libfqfft: C++ library for FFTs in Finite Fields.",
  year = "2018",
  url = "https://github.com/scipr-lab/libfqfft"
}

@article{gabizon2019plonk,
  title={Plonk: Permutations over lagrange-bases for oecumenical noninteractive arguments of knowledge},
  author={Gabizon, Ariel and Williamson, Zachary J and Ciobotaru, Oana},
  journal={Cryptology ePrint Archive},
  year={2019}
}

@misc{ingonyama_2022, title={Hardware-friendliness of HyperPlonk}, howpublished={\href{https://medium.com/@ingonyama/hardware-friendliness-of-hyperplonk-491d8c86605}{https://medium.com/@ingonyama/hardware-friendliness-of-hyperplonk-491d8c86605}}, journal={Medium}, author={Ingonyama}, year={2022}, month={Dec} }

@misc{irreducible_2023, title={A Throughput-Optimized FPGA Architecture for Goldilocks NTT - Irreducible}, howpublished={\href{https://www.irreducible.com/posts/fpga-architecture-for-goldilocks-ntt}{https://www.irreducible.com/posts/fpga-architecture-for-goldilocks-ntt}}, journal={Irreducible.com}, author={Irreducible, Team}, year={2023} }

@incollection{franchetti2011fft,
  title={FFT (fast fourier transform)},
  author={Franchetti, Franz and P{\"u}schel, Markus},
  booktitle={Encyclopedia of Parallel Computing},
  pages={658--671},
  year={2011},
  publisher={Springer}
}

@INPROCEEDINGS{rpu,
  author={Soni, Deepraj and Neda, Negar and Zhang, Naifeng and Reynwar, Benedict and Gamil, Homer and Heyman, Benjamin and Nabeel, Mohammed and Badawi, Ahmad Al and Polyakov, Yuriy and Canida, Kellie and Pedram, Massoud and Maniatakos, Michail and Cousins, David Bruce and Franchetti, Franz and French, Matthew and Schmidt, Andrew and Reagen, Brandon},
  booktitle={2023 IEEE International Symposium on Performance Analysis of Systems and Software (ISPASS)}, 
  title={RPU: The Ring Processing Unit}, 
  year={2023},
  volume={},
  number={},
  pages={272-282},
  doi={10.1109/ISPASS57527.2023.00034}}

@INPROCEEDINGS{ciflow,
  author={Neda, Negar and Ebel, Austin and Reynwar, Benedict and Reagen, Brandon},
  booktitle={2024 IEEE International Symposium on Performance Analysis of Systems and Software (ISPASS)}, 
  title={CiFlow: Dataflow Analysis and Optimization of Key Switching for Homomorphic Encryption}, 
  year={2024},
  volume={},
  number={},
  pages={61-72},
  doi={10.1109/ISPASS61541.2024.00016}}

@inproceedings{karthik,
author = {Garimella, Karthik and Ghodsi, Zahra and Jha, Nandan Kumar and Garg, Siddharth and Reagen, Brandon},
title = {Characterizing and Optimizing End-to-End Systems for Private Inference},
year = {2023},
isbn = {9781450399180},
publisher = {Association for Computing Machinery},
address = {New York, NY, USA},
url = {https://doi.org/10.1145/3582016.3582065},
doi = {10.1145/3582016.3582065},
booktitle = {Proceedings of the 28th ACM International Conference on Architectural Support for Programming Languages and Operating Systems, Volume 3},
pages = {89–104},
numpages = {16},
location = {Vancouver, BC, Canada},
series = {ASPLOS 2023}
}

@INPROCEEDINGS {legozk,
author = { Yang, Zhengbang and Zhao, Lutan and Li, Peinan and Liu, Han and Li, Kai and Zhao, Boyan and Meng, Dan and Hou, Rui },
booktitle = { 2025 IEEE International Symposium on High Performance Computer Architecture (HPCA) },
title = {{ LegoZK: A Dynamically Reconfigurable Accelerator for Zero-Knowledge Proof }},
year = {2025},
volume = {},
ISSN = {},
pages = {113-126},
doi = {10.1109/HPCA61900.2025.00020},
url = {https://doi.ieeecomputersociety.org/10.1109/HPCA61900.2025.00020},
publisher = {IEEE Computer Society},
address = {Los Alamitos, CA, USA},
month =mar}

@inproceedings{unizk,
author = {Wang, Cheng and Gao, Mingyu},
title = {UniZK: Accelerating Zero-Knowledge Proof with Unified Hardware and Flexible Kernel Mapping},
year = {2025},
isbn = {9798400706981},
publisher = {Association for Computing Machinery},
address = {New York, NY, USA},
url = {https://doi.org/10.1145/3669940.3707228},
doi = {10.1145/3669940.3707228},
booktitle = {Proceedings of the 30th ACM International Conference on Architectural Support for Programming Languages and Operating Systems, Volume 1},
pages = {1101–1117},
numpages = {17},
location = {Rotterdam, Netherlands},
series = {ASPLOS '25}
}

@inproceedings{ben2019aurora,
  title={Aurora: Transparent succinct arguments for R1CS},
  author={Ben-Sasson, Eli and Chiesa, Alessandro and Riabzev, Michael and Spooner, Nicholas and Virza, Madars and Ward, Nicholas P},
  booktitle={Annual international conference on the theory and applications of cryptographic techniques},
  pages={103--128},
  year={2019},
  organization={Springer}
}

@misc{janeStreet_2022, title={Accelerating zk-SNARKs - MSM and NTT algorithms on FPGAs with Hardcaml}, howpublished={\href{https://blog.janestreet.com/zero-knowledge-fpgas-hardcaml/}{https://blog.janestreet.com/zero-knowledge-fpgas-hardcaml/}}, journal={Jane Street Blog}, author={Ray, Andrew and Devlin, Benjamin and Quah, Fu Yong and Yesantharao, Rahul}, year={2022} }

@misc{intel_xeon, title={Intel® Xeon® Gold 5218 Processor (22M Cache, 2.30 GHz)}, 
howpublished={\href{https://www.intel.com/content/www/us/en/products/sku/192444/intel-xeon-gold-5218-processor-22m-cache-2-30-ghz/specifications.html}{https://www.intel.com/content/www/us/en/products/
                    sku/192444/intel-xeon-gold-5218-processor-22m-cache-2-30-ghz/specifications.html}}, journal={Intel}, author={Intel}, year={2025} }

@book{thaler_proofs_args_zk,
  author       = {Justin Thaler},
  title        = {Proofs, Arguments, and Zero-Knowledge},
  year         = {2022},
  url          = {https://people.cs.georgetown.edu/jthaler/ProofsArgsAndZK.pdf},
}

@inproceedings{ligero_pcs,
    author = {Ames, Scott and Hazay, Carmit and Ishai, Yuval and Venkitasubramaniam, Muthuramakrishnan},
    title = {Ligero: Lightweight Sublinear Arguments Without a Trusted Setup},
    year = {2017},
    isbn = {9781450349468},
    publisher = {Association for Computing Machinery},
    address = {New York, NY, USA},
    url = {https://doi.org/10.1145/3133956.3134104},
    doi = {10.1145/3133956.3134104},
    booktitle = {Proceedings of the 2017 ACM SIGSAC Conference on Computer and Communications Security},
    pages = {2087–2104},
    numpages = {18},
    location = {Dallas, Texas, USA},
    series = {CCS '17}
}

@inproceedings{zk_proof_gmr,
    author = {Goldwasser, S and Micali, S and Rackoff, C},
    title = {The knowledge complexity of interactive proof-systems},
    year = {1985},
    isbn = {0897911512},
    publisher = {Association for Computing Machinery},
    address = {New York, NY, USA},
    url = {https://doi.org/10.1145/22145.22178},
    doi = {10.1145/22145.22178},
    booktitle = {Proceedings of the Seventeenth Annual ACM Symposium on Theory of Computing},
    pages = {291–304},
    numpages = {14},
    location = {Providence, Rhode Island, USA},
    series = {STOC '85}
}

@inproceedings{zksnarks_bcct12,
    author = {Bitansky, Nir and Canetti, Ran and Chiesa, Alessandro and Tromer, Eran},
    title = {From extractable collision resistance to succinct non-interactive arguments of knowledge, and back again},
    year = {2012},
    isbn = {9781450311151},
    publisher = {Association for Computing Machinery},
    address = {New York, NY, USA},
    url = {https://doi.org/10.1145/2090236.2090263},
    doi = {10.1145/2090236.2090263},
    booktitle = {Proceedings of the 3rd Innovations in Theoretical Computer Science Conference},
    pages = {326–349},
    numpages = {24},
    location = {Cambridge, Massachusetts},
    series = {ITCS '12}
}

@inproceedings{iops_bcs,
    author = {Ben-Sasson, Eli and Chiesa, Alessandro and Spooner, Nicholas},
    title = {Interactive Oracle Proofs},
    year = {2016},
    isbn = {9783662536438},
    publisher = {Springer-Verlag},
    address = {Berlin, Heidelberg},
    url = {https://doi.org/10.1007/978-3-662-53644-5_2},
    doi = {10.1007/978-3-662-53644-5_2},
    booktitle = {Proceedings, Part II, of the 14th International Conference on Theory of Cryptography - Volume 9986},
    pages = {31–60},
    numpages = {30}
}

@inproceedings{fiat1986prove,
  title={How to Prove Yourself: Practical Solutions to Identification and Signature Problems},
  author={Fiat, Amos and Shamir, Adi},
  booktitle={Advances in Cryptology – CRYPTO '86},
  series={Lecture Notes in Computer Science},
  volume={263},
  pages={186--194},
  year={1986},
  publisher={Springer},
  doi={10.1007/3-540-47721-7_12}
}

@article{rothblum2024note,
  title={A note on efficient computation of the multilinear extension},
  author={Rothblum, Ron D},
  journal={Cryptology ePrint Archive},
  year={2024}
}

@article{dao2024more,
  title={More optimizations to sum-check proving},
  author={Dao, Quang and Thaler, Justin},
  journal={Cryptology ePrint Archive},
  year={2024}
}

@article{dao2024constraint,
  title={Constraint-packing and the sum-check protocol over binary tower fields},
  author={Dao, Quang and Thaler, Justin},
  journal={Cryptology ePrint Archive},
  year={2024}
}

@article{bagad2025speeding,
  title={Speeding Up Sum-Check Proving},
  author={Bagad, Suyash and Dao, Quang and Domb, Yuval and Thaler, Justin},
  journal={Cryptology ePrint Archive},
  year={2025}
}

@inproceedings{chiesa2020marlin,
  title={Marlin: Preprocessing zkSNARKs with universal and updatable SRS},
  author={Chiesa, Alessandro and Hu, Yuncong and Maller, Mary and Mishra, Pratyush and Vesely, Noah and Ward, Nicholas},
  booktitle={Annual International Conference on the Theory and Applications of Cryptographic Techniques},
  pages={738--768},
  year={2020},
  organization={Springer}
}

@inproceedings{moma_micro,
author = {Zhang, Naifeng and Fu, Sophia and Franchetti, Franz},
title = {Towards Closing the Performance Gap for Cryptographic Kernels Between CPUs and Specialized Hardware},
year = {2025},
isbn = {9798400715730},
publisher = {Association for Computing Machinery},
address = {New York, NY, USA},
url = {https://doi.org/10.1145/3725843.3756120},
doi = {10.1145/3725843.3756120},
booktitle = {Proceedings of the 58th IEEE/ACM International Symposium on Microarchitecture},
pages = {1704–1718},
numpages = {15},
location = {
},
series = {MICRO '25}
}

@inproceedings{moma_cgo,
author = {Zhang, Naifeng and Franchetti, Franz},
title = {Code Generation for Cryptographic Kernels using Multi-word Modular Arithmetic on GPU},
year = {2025},
isbn = {9798400712753},
publisher = {Association for Computing Machinery},
address = {New York, NY, USA},
url = {https://doi.org/10.1145/3696443.3708948},
doi = {10.1145/3696443.3708948},
booktitle = {Proceedings of the 23rd ACM/IEEE International Symposium on Code Generation and Optimization},
pages = {476–492},
numpages = {17},
location = {Las Vegas, NV, USA},
series = {CGO '25}
}

@inproceedings{zkgpt,
author = {Qu, Wenjie and Sun, Yijun and Liu, Xuanming and Lu, Tao and Guo, Yanpei and Chen, Kai and Zhang, Jiaheng},
title = {zkGPT: an efficient non-interactive zero-knowledge proof framework for LLM inference},
year = {2025},
isbn = {978-1-939133-52-6},
publisher = {USENIX Association},
address = {USA},
booktitle = {Proceedings of the 34th USENIX Conference on Security Symposium},
articleno = {106},
numpages = {19},
location = {Seattle, WA, USA},
series = {SEC '25}
}

@inproceedings{zkcnn,
author = {Liu, Tianyi and Xie, Xiang and Zhang, Yupeng},
title = {zkCNN: Zero Knowledge Proofs for Convolutional Neural Network Predictions and Accuracy},
year = {2021},
isbn = {9781450384544},
publisher = {Association for Computing Machinery},
address = {New York, NY, USA},
url = {https://doi.org/10.1145/3460120.3485379},
doi = {10.1145/3460120.3485379},
booktitle = {Proceedings of the 2021 ACM SIGSAC Conference on Computer and Communications Security},
pages = {2968–2985},
numpages = {18},
location = {Virtual Event, Republic of Korea},
series = {CCS '21}
}

@inproceedings{zkml,
author = {Chen, Bing-Jyue and Waiwitlikhit, Suppakit and Stoica, Ion and Kang, Daniel},
title = {ZKML: An Optimizing System for ML Inference in Zero-Knowledge Proofs},
year = {2024},
isbn = {9798400704376},
publisher = {Association for Computing Machinery},
address = {New York, NY, USA},
url = {https://doi.org/10.1145/3627703.3650088},
doi = {10.1145/3627703.3650088},
booktitle = {Proceedings of the Nineteenth European Conference on Computer Systems},
pages = {560–574},
numpages = {15},
location = {Athens, Greece},
series = {EuroSys '24}
}

@inproceedings{zkllm,
author = {Sun, Haochen and Li, Jason and Zhang, Hongyang},
title = {zkLLM: Zero Knowledge Proofs for Large Language Models},
year = {2024},
isbn = {9798400706363},
publisher = {Association for Computing Machinery},
address = {New York, NY, USA},
url = {https://doi.org/10.1145/3658644.3670334},
doi = {10.1145/3658644.3670334},
booktitle = {Proceedings of the 2024 on ACM SIGSAC Conference on Computer and Communications Security},
pages = {4405–4419},
numpages = {15},
location = {Salt Lake City, UT, USA},
series = {CCS '24}
}

@inproceedings{mo2023towards,
  title={Towards fast and scalable private inference},
  author={Mo, Jianqiao and Garimella, Karthik and Neda, Negar and Ebel, Austin and Reagen, Brandon},
  booktitle={Proceedings of the 20th ACM International Conference on Computing Frontiers},
  pages={322--328},
  year={2023}
}

@article{dhyani2024privit,
  title={PriViT: Vision Transformers for Private Inference},
  author={Dhyani, Naren and Mo, Jianqiao Cambridge and Yubeaton, Patrick and Cho, Minsu and Joshi, Ameya and Garg, Siddharth and Reagen, Brandon and Hegde, Chinmay},
  journal={Transactions on Machine Learning Research},
  year={2024}
}

@article{geng2026ppimce,
  title={PPIMCE: In-Memory Computing Fabric for Privacy Preserving Computing},
  author={Geng, Haoran and Mo, Jianqiao and Reis, Dayane and Takeshita, Jonathan and Jung, Taeho and Reagen, Brandon and Niemier, Michael and Hu, Xiaobo Sharon},
  journal={Journal of Computer Science and Technology},
  volume={41},
  number={1},
  pages={83--102},
  year={2026},
  publisher={Springer}
}

@inproceedings{kim2025anaheim,
  title={Anaheim: Architecture and Algorithms for Processing Fully Homomorphic Encryption in Memory},
  author={Kim, Jongmin and Yun, Sungmin and Ji, Hyesung and Choi, Wonseok and Kim, Sangpyo and Ahn, Jung Ho},
  booktitle={2025 IEEE International Symposium on High Performance Computer Architecture (HPCA)},
  pages={1158--1173},
  year={2025},
  organization={IEEE}
}

@inproceedings{cho2024apint,
  title={APINT: A Full-Stack Framework for Acceleration of Privacy-Preserving Inference of Transformers based on Garbled Circuits},
  author={Cho, Hyunjun and Jeon, Jaeho and Heo, Jaehoon and Kim, Joo-Young},
  booktitle={Proceedings of the 43rd IEEE/ACM International Conference on Computer-Aided Design},
  pages={1--9},
  year={2024}
}

@inproceedings{zhu2026efficient,
  title={An Efficient and Scalable Hardware Architecture for Number Theoretic Transform on FPGA with Design Automation},
  author={Zhu, Yilan and Yang, Geng and Tian, Xingyu and Kumarathunga, Dilshan and Kong, Liang and Deng, Xianglong and Fan, Shengyu and Fan, Guang and Shi, Guiming and Chen, Lei and others},
  booktitle={2026 IEEE International Symposium on High Performance Computer Architecture (HPCA)},
  pages={1--14},
  year={2026},
  organization={IEEE}
}
